 \documentclass[final,3p,times]{elsarticle}

\usepackage{amssymb}
\usepackage{amsmath}
\usepackage{graphicx}%
\usepackage{multirow}%
\usepackage{amssymb}
\usepackage{amsfonts}%
\usepackage{amsthm}%
\usepackage{mathrsfs}%
\usepackage[title]{appendix}%
\usepackage{xcolor}%
\usepackage{textcomp}%
\usepackage{manyfoot}%
\usepackage{booktabs}%
\usepackage{algorithm}%
\usepackage{algorithmicx}%
\usepackage{algpseudocode}%
\usepackage{listings}%
\usepackage{subcaption}
\usepackage{tabularx}
\usepackage{makecell}
\usepackage{lscape}
\usepackage{threeparttable}
\usepackage{caption}
\usepackage[figuresright]{rotating}
\usepackage{color, soul, xcolor}
\usepackage{pifont}
\usepackage{nomencl}
\usepackage{etoolbox}

\journal{arXiv}

\begin{document}

\begin{frontmatter}



\title{Generalizability of local neural operator: example for elastodynamic problems}

\author[label1,label2]{Hongyu Li}
\ead{lihongyu_0@163.com}

\author[label1]{Ximeng Ye}
\ead{yeximeng@stu.xjtu.edu.cn}

\author[label1]{Lei He}
\ead{leiluodelei@nudt.edu.cn}

\author[label1]{Weiqi Qian\corref{corrauthor}}
\ead{qwqhyy@sina.com}

\author[label2]{Peng Jiang}
\ead{jiangpeng219@xjtu.edu.cn}

\author[label2]{Tiejun Wang\corref{corrauthor}}
\ead{wangtj@xjtu.edu.cn}

\cortext[corrauthor]{Corresponding authors}

\affiliation[label1]{organization={China Aerodynamics Research and Development Center},
            city={Mianyang},
            postcode={621000},
            state={Sichuan},
            country={China}}

\affiliation[label2]{organization={State Key Laboratory for Strength and Vibration of Mechanical Structures, Department of Engineering Mechanics, Xi’an Jiaotong University},
            city={Xi'an},
            postcode={710049},
            state={Shaanxi},
            country={China}}

\begin{abstract}
Local neural operator (LNO) conception has provided a feasible way for scientific computations. 
The LNO learns transient partial differential equations from random field samples, and then the pre-trained LNO solves practical problems on specific computational domains. 
For applications, we may ask: \textit{Are the training samples rich enough? To what extent can we trust the solutions obtained from pre-trained LNO models for unknown cases?} 
The generalizability of LNO could answer these questions. 
Here, we propose to use two plain scalar features, the amplitude and wavenumber of the input functions, to indicate the richness of training samples and to evaluate the generalization error of pre-trained LNO. 
In elastodynamic practices, we find that isolated evolving wavenumber modes for Lamé-Navier equation caused the training dataset to lack mode diversity. 
By data supplementation and model fine-tuning targeting to the discovered lack modes, the pre-trained and fine-tuned LNO model solves Lamb problem correctly and efficiently. 
These results and the proposed generalization criteria provide a paradigm for LNO applications.

\end{abstract}




\end{frontmatter}

\setlength{\nomlabelwidth}{80pt}
\setlength{\nomitemsep}{-1.35mm}
\renewcommand\nomgroup[1]{%
  \item[\bfseries
  \ifstrequal{#1}{A}{\textit{For local neural operator conception}}{%
  \ifstrequal{#1}{B}{\textit{For generalizability evaluation and analysis}}{%
  \ifstrequal{#1}{C}{\textit{For application demos and boundary treatment}}{
  \ifstrequal{#1}{D}{\!\textit{For Lamé-Navier equation and LNO modeling analyses}}
  }}}
]}
\makenomenclature

\nomenclature[A,01]{$\mathcal{G}_\mathrm{L}$}{Target local-related time-marching operator for dynamics of continuum systems}
\nomenclature[A,02]{$\mathcal{G}_\theta$}{LNO for approximating $\mathcal{G}_\mathrm{L}$ with the set $\theta$ of trainable weights}
\nomenclature[A,03]{$\boldsymbol{u}_t\left(\boldsymbol{x}\right)$}{Physical fields at time $t$ (displacement for elastodynamics)}
\nomenclature[A,04]{$\dot{\boldsymbol{u}}\left(\boldsymbol{x}\right),\ddot{\boldsymbol{u}}\left(\boldsymbol{x}\right)$}{Time derivatives of the physical fields (velocity and acceleration for elastodynamics)}
\nomenclature[A,05]{$\tilde{\boldsymbol{u}}_{t+\Delta t}\left(\boldsymbol{x}\right)$}{Predicted physical fields at time $t+\Delta t$ by LNO}
\nomenclature[A,06]{$D_1,D_2$}{Unit output/input domain}
\nomenclature[A,07]{$r_\mathrm{min}$}{Local-related range}
\nomenclature[A,08]{$\Omega$}{Computational domain of the model problems}
\nomenclature[A,09]{$\Omega_\mathrm{in},\Omega_\mathrm{out}$}{Complete output/input domain for LNO}
\nomenclature[A,10]{$\lambda,\mu$}{Lamé coefficients for elastodynamic computations}
\nomenclature[A,11]{$\rho$}{Density for elastodynamic computations}
\nomenclature[A,12]{$\Delta x$}{Grid size for discretization}
\nomenclature[A,13]{$N_t,N_s$}{Number of discretized points in time and space\vspace{10pt}}

\nomenclature[B,01]{$\mathcal{E}$}{Integrated mean error}
\nomenclature[B,02]{$\mathcal{E}_\mathrm{gen}$}{Generalization error (integrated mean)}
\nomenclature[B,03]{$\mathcal{P}$}{Data distribution}
\nomenclature[B,04]{$\left(X,Y\right),\,\left(X_i,Y_i\right)$}{Random variable of input and output pairs and the $i$-th sampled pair}
\nomenclature[B,05]{$\mathcal{F}_X$}{A feature vector for representing function $X$}
\nomenclature[B,06]{$\boldsymbol{B}\left(x\right)$}{Function basis for random data generation}
\nomenclature[B,07]{$\boldsymbol{\Lambda},\lambda_{ij}$}{Random square martrix and its components}
\nomenclature[B,08]{$a$}{Amplitude coefficient for data functions}
\nomenclature[B,09]{$k_1,k_2,k$}{Wavenumber coefficient for data functions, in the two directions and their mean}
\nomenclature[B,10]{$\phi$}{Scalar potential function}
\nomenclature[B,11]{$\boldsymbol{\psi}$}{Nondivergent vector potential function}
\nomenclature[B,12]{$\phi_0,\phi_1$}{Initial condition for the example scalar problem}
\nomenclature[B,13]{$B\left(\boldsymbol{x},t\right)$}{A ball of radius $t$ centered at $\boldsymbol{x}$}
\nomenclature[B,14]{$\hat{\phi},\hat{\phi_0},\hat{\phi_1}$}{Fourier componets of $\phi,\phi_0,\phi_1$\vspace{10pt}}

\nomenclature[C,01]{$\Omega_1,\Omega_2,\Omega_3$}{The extended and divided computational domains for LNO prediction}
\nomenclature[C,02]{$\boldsymbol{u}^*_t,\dot{\boldsymbol{u}}^*_t$}{Virtual fields at time $t$ on extended domain along physical boundaries}
\nomenclature[C,03]{$S\left(t\right)$}{Transient source function}
\nomenclature[C,04]{$A$}{Amplitude coefficient for point source}
\nomenclature[C,05]{$f$}{Frequency of Ricker’s wavelet point source}
\nomenclature[C,06]{$t_0$}{Ending time of point source}
\nomenclature[C,07]{$\boldsymbol{n}$}{Normal vector outword the boundary}
\nomenclature[C,08]{$\boldsymbol{\sigma}\left(\boldsymbol{x}\right),\sigma_{ij}$}{Stress field and its component}
\nomenclature[C,09]{$b\left(x_1\right)=x_2$}{Equation for curved boundary}
\nomenclature[C,10]{$\boldsymbol{x}_u,\boldsymbol{x}_\sigma$}{Points on displacement and stress boundary}
\nomenclature[C,11]{$\mathcal{L}_u,\mathcal{L}_\sigma$}{Optimization loss for displacement and stress boundary imposing}
\nomenclature[C,12]{$\bar{\boldsymbol{u}},\dot{\bar{\boldsymbol{u}}},\bar{\boldsymbol{\sigma}}$}{Boundary displacement, velocity, and stress values\vspace{10pt}}

\nomenclature[D,01]{$V$}{Computational domain}
\nomenclature[D,02]{$S$}{Boundary}
\nomenclature[D,021]{$\boldsymbol{f},f_i$}{Body force and its components}
\nomenclature[D,03]{$v_\mathrm{p},v_\mathrm{s}$}{Propagating speeds of P-wave and S-wave}
\nomenclature[D,04]{$\boldsymbol{G},G_{in}$}{Green’s function and its components}
\nomenclature[D,05]{$C_{ijpq}$}{Components in a 4-th order tensor of elastic coefficients}
\nomenclature[D,06]{$t_\mathrm{p},t_\mathrm{s}$}{Arrival time of P-wave and S-wave}
\nomenclature[D,07]{$\gamma_n$}{Direction coefficient for direction $n$}
\nomenclature[D,08]{$\boldsymbol{T},T_i$}{Inclined plane stress on boundaries and its components}

\printnomenclature
\emph{\textbf{Remark}: Italic symbols are variables, bold symbols are vectors or higher-order tensors. }

\section{Introduction}\label{sec1}

Operator learning are now forging toward a possible revolution for scientific computation \cite{Higgins2021,Zhang2023}. 
Starting from deep operator network (DeepONet) \cite{LuLu2021}, neural operator (NO) \cite{Li2020c}, and Fourier neural operator (FNO) \cite{LiZongyi2020}, the consensus we met on the definition of operator learning is that, neural networks learn to approximate mappings between function spaces (the input and output are continuous functions, or vectors with infinite dimensions) \cite{Kovachki2021}. 
Recent progress in operator learning methods is primarily made in solving partial differential equations (PDE) for scientific computation or engineering design \cite{Azizzadenesheli2023a}, and it is growing rapidly to great prosperity. 
Typical progresses are made in neural operator architecture \cite{He2023,Hao2023b,Kissas2022}, kernel design for neural operators \cite{Gupta2021a,Fanaskov2023,Navaneeth2024}, learning on non-plain manifolds and complex geometries \cite{CHEN2024,Li2023a}, learning for dynamics systems \cite{He2024,Xiong2023}, as well as reports of advanced applications in weather forecast \cite{PathakJ2022}, dynamics of fluids and heat transfer \cite{Ye2023,Koric2023a}, wave propagation and seismology \cite{ClarkDiLeoni2023,Haghighat2024,YangY2021}, materials \cite{Oommen2022,Li2023}, etc. 
We can summarize that a primary advantage of operator learning methods for solving PDE is the \textit{superior efficiency} obtained from \textit{the reuse of pre-trained neural operator models}. 

\begin{figure}[ht]
\centering
\includegraphics[width=0.9\textwidth]{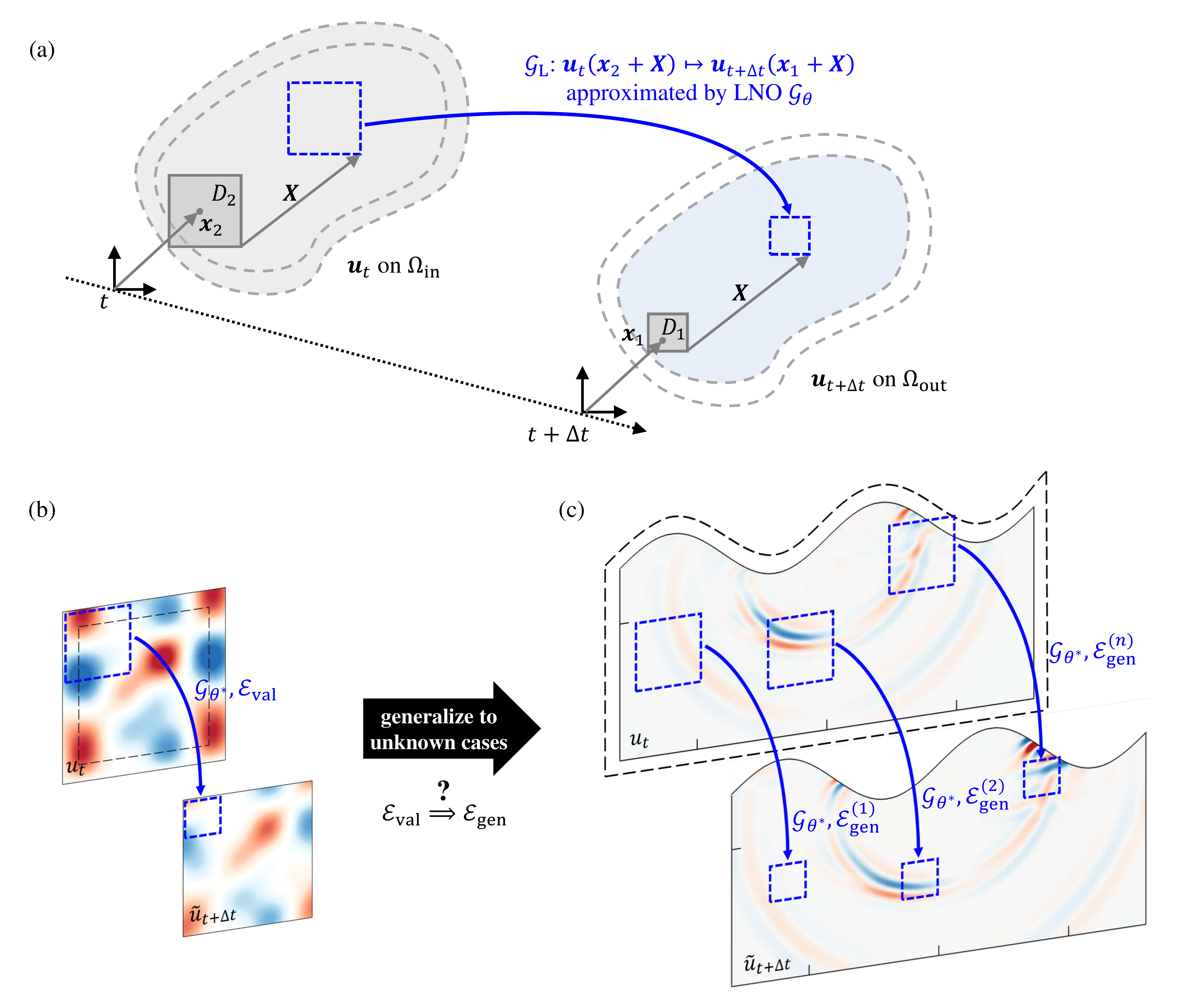}
\caption{The generalizing process to use pre-trained LNO models. 
(a) LNO conception enables variable domains for input functions on $\Omega_{\rm{in}}$ and output functions on $\Omega_{\rm{out}}$. 
(b) The pre-trained LNO $\mathcal{G}_{\theta^* }$ and its validation error $\mathcal{E}_{\rm{gen}}$. 
(c) The pre-trained LNO $\mathcal{G}_{\theta^* }$ can have diverse error $\mathcal{E}_{\rm{gen}}^{(n)}$ on different local parts when solving unknown cases.}
\label{fig:LNOconcept}
\end{figure}

The conception of local neural operator (LNO) \cite{LiHongyu2022} amplifies this advantage by extending the reusable scope of pre-trained neural operator models. 
When learning transient PDE representing dynamics of continuum systems in physics, it separates case-specific initial and boundary conditions from the governing equation by introducing a local-related condition and a shift-invariant condition as in Fig.~\ref{fig:LNOconcept}(a), thus making one pre-trained LNO model to solely encode the governing equation, and being capable of solving practical problems on varied computational domains. 
This local-related idea coincides with both early conventional compact numerical schemes \cite{Deng2000,Zhang2008,E1996} and later discussions on operator learning \cite{Liu-Schiaffini2024}. 
Beyond the amplified efficiency advantage, the reliability of the solutions becomes the next issue for LNO toward scientific and engineering applications, i.e., we have to answer: \textit{To what extent can we trust the solutions obtained from pre-trained LNO models for unknown cases?}
This reliability issue can be technically described as a pre-evaluation of the solving error, which should be closely related to the training and validation error for statistical learning methods such as the present operator learning method, the LNO. 
Specifically, the LNO model~\cite{LiHongyu2022} is trained on randomly generated samples on boundary-free domains, as shown in Fig.~\ref{fig:LNOconcept}(b), then, it can be used to solve on complex domains as in Fig.~\ref{fig:LNOconcept}(c). 
In this case, what we want is to find out the relation of validation error $\mathcal{E}_{\rm{val}}$ and generalization error $\mathcal{E}_{\rm{gen}}$. 
We are particularly interested in the condition and the changing rules for how $\mathcal{E}_{\rm{gen}}$ can be inferred from $\mathcal{E}_{\rm{val}}$.

Evaluating the generalization error (also called extrapolation error) of neural operators is troubled by both the complex diverse learning tasks and varied neural network architectures.
About the learning tasks, a discussion about the extrapolation of DeepONet found that different equations on different shapes of computational domains, or different parameters used in Gaussian random fields for generating field samples, can lead to diverse extrapolation rules \cite{Zhu2023}. 
Even for the present work that the governing equation is fixed, different local fields can have diverse errors as in Fig.~\ref{fig:LNOconcept}(c). 
About the neural network architecture for approximating the operators, it is revealed that, neural networks with linear activation functions such as ReLU can hardly extrapolate nonlinear functions, and the successful extrapolation relies on proper encoding of task-specific features \cite{Xu2021}. 
Besides, the successful generalization (extrapolation) of LNO to varied domains \cite{LiHongyu2022} could be an example which neatly utilizes the local-related and shift-invariant nature of transient PDEs. 
These works show that, the generalizability of learning methods should be case-specific studied as the learning task, the neural network architecture, and their interactions are complex. 
Technically, there are two tough issues regarding the generalizability of pre-trained LNOs. 
One is the infinite input and output spaces, which are hard to describe and parameterize. 
The other is that, the error of learning methods is usually discussed by distributions, yet the pre-trained LNO solves practical problems by instances. 
Though there are uncertainty estimation theories such as the Bayesian analysis or conformal prediction, specifically applied to DeepONet \cite{Lin2023,Moya2023,Moya2025}, practical applications are still troubled by the independent and identically distributed (\textit{i.i.d.}) assumption or large computing consumptions.

This work tries a direct way to evaluate the errors of solutions obtained by pre-trained local neural operators. 
The generalization error is defined regarding the primary scalar features of the input functions, thus enabling a practical and continuous evaluation metric regarding scalar extrapolation distances, which can be used to delimit a safe range for PDE-solving applications. 
This generalization metric revealed and improved the generalization shortage of pre-trained LNO to diverse wavenumbers in learning the elastodynamic equation. 
These discoveries and advances could make LNO a fast tool for large-scale analyses, which field has also suffered from shortages of computer resources \cite{Fu2017}. 
This paper is organized as follows. 
Section~\ref{sec2} introduces preliminaries of LNO, model equations, the baseline training and validation. 
In section~\ref{sec3}, we define generalization error metrics, then, the generalizability of baseline pre-trained LNO models in learning the considered transient equations is evaluated and discussed.
Section~\ref{sec4} focuses on the generalizability of pre-trained LNO to diverse wavenumber modes. 
By data supplementation and fine-tuning following these analyses, the generalizability is further improved.
Moreover, elastodynamic application demos are presented in section~\ref{sec5}, several benchmark model problems are solved by the pre-trained LNO and compared to conventional solvers, showing positive potential for more applications. 
Section~\ref{sec6} concludes this paper.

\section{Preliminaries}\label{sec2}

\subsection{Learn dynamics of continuum systems with LNO\label{sec2:1}}

LNO approximates a local-related and shift-invariant time-marching operator for transient partial differential equations \cite{LiHongyu2022}. 
In applied mechanics, this represents dynamics of continuum systems. 
The target operator $\mathcal{G}_{\rm{L}}$ for LNO can be written as
\begin{equation}
\begin{split}
\mathcal{G}_{\rm{L}}:\boldsymbol{u}_t(\boldsymbol{x}_2^\prime) &\mapsto \boldsymbol{u}_{t+\Delta t}(\boldsymbol{x}^\prime_1), \qquad t\geq 0,\boldsymbol{x}^\prime_1\in \Omega_{\rm{out}},\boldsymbol{x}_2^\prime\in \Omega_{\rm{in}}, \\
\Omega_{\rm{out}}&=\left\{\boldsymbol{x}_1+\boldsymbol{X}\ \middle|\ \boldsymbol{x}_1\in D_1,\boldsymbol{X}\in\mathcal{X}\right\},\\
\Omega_{\rm{in}}&=\left\{\boldsymbol{x}_2+\boldsymbol{X}\ \middle|{\ \boldsymbol{x}}_2\in D_2,\boldsymbol{X}\in\mathcal{X}\right\}.
\label{eq:1}
\end{split}
\end{equation}
The input and output functions $\boldsymbol{u}_t$ on $\Omega_{\rm{in}}$ and $\boldsymbol{u}_{t+\Delta t}$ on $\Omega_{\rm{out}}$ are respectively physical fields for the considered equation at time level $t$ and $t+\Delta t$. $D_1$ and $D_2$ are respectively the unit output and input domain. They fulfill the local related condition as
\begin{equation}
\frac{\partial\boldsymbol{u}_{t+\Delta t}(\boldsymbol{x}_1)}{\partial\boldsymbol{u}_{t}(\boldsymbol{x}_2)}=0,\qquad\forall\,\lVert\boldsymbol{x}_{1}-\boldsymbol{x}_{2}\rVert>r,\;\boldsymbol{x}_1\in D_1,\;\boldsymbol{x}_2\in D_2
\label{eq:2}
\end{equation}
$r$ is a finite positive number. 
Its minimum $r_{\rm{min}}$ represent the maximal related range, which is a primary feature of LNO \cite{Ye2023}. 
$\mathcal{X}$ in Eq.~(\ref{eq:1}) is the set of shifting vectors $\boldsymbol{X}$, which can be arbitrarily given according to specific problem to be solved. 
It thus makes the solving domain $\Omega_{\rm{in}}$ and $\Omega_{\rm{out}}$ variable for one trained LNO model. 
The following introduces the LNO architecture \cite{Ye2023} used for the present study. 
The network follows a lifting-projection frame and owns four two-path blocks as the main body. 
The two paths in blocks are physical path and spectral path, which respectively transforms the functions directly in physical space or in a spectral space constructed by Legendre polynomials. 
All the operations used in the architecture should be local-related for ensuring a local-related overall architecture of LNO. 
This study adopts architecture hyperparameter as, the number of blocks $n=4$, the size of local window for spectral transform $N=12$, the number of reserved spectral components $M=6$, and the number of repetitions in geometric decomposition $k=2$.

The data samples for training and validation are generated as follows \cite{LiHongyu2022}. 
Firstly, define a model problem on square domains with circular boundary conditions and random initial conditions. 
We use simple trigonometric basis with random coefficients for initial conditions. 
For 2D cases with coordinates $x,\,y$, it is written as
\begin{equation}
u_0\left(x,y\right)=\boldsymbol{B}^T\left(x\right)\mathbf{\Lambda}\boldsymbol{B}\left(y\right),\ \ x\in\left[1,-1\right],y\in\left[1,-1\right]
\label{eq:3}
\end{equation}
where $\boldsymbol{B}\left(x\right)=\left\{\sin{\pi x},\sin{2\pi x},\cos{\pi x},\cos{2\pi x}\right\}^T$, $\boldsymbol{\Lambda}=\left\{\lambda_{ij}\right\}\, (i,j=1,2,3,4)$ and $\lambda_{ij}\sim N\left(0,1\right)$. 
Then, solve the model problems by conventional numeric solver, specifically the finite element method (FEM). 
Finally, the solutions are recorded in field series. Supervised by bunches of these field data, LNO are trained following a recurrent scheme.

\subsection{Model equations\label{sec2:2}}

The scope of LNO includes hyperbolic and parabolic transient partial differential equations, thereby fulfilling or approximately fulfilling the local-related condition in Eq.~(\ref{eq:2}). This work considers two typical model equations from applied mechanics. One is the viscous Burgers equation which owns first-order hyperbolic and parabolic features. The other is Lamé-Navier equation which is second-order hyperbolic. We choose these two model equations to representationally study the generalizability of LNO in convection-diffusion systems (fluids) and waving systems (solids). 

Viscous Burgers equation has been widely used as model problem in neural operator researches \cite{Xiong2023,Burgers1948}. For 2D cases, it is 
\begin{equation}
\dot{\boldsymbol{v}}=\nu\Delta\boldsymbol{v}-\boldsymbol{v}\cdot\nabla\boldsymbol{v}.
\label{eq:4}
\end{equation}
$\boldsymbol{v}\left(\boldsymbol{x},t\right)=\left\{v_1,v_2\right\}\left(\boldsymbol{x},t\right), \boldsymbol{x}\in\Omega\subset\mathbb{R}^2,t>0,$ is the velocity to be solved, $\nu$ is the viscosity. LNO learns this equation by approximating its time-marching operator as 
\begin{equation}
\mathcal{G}_{\rm{L}}^{\rm{\left(Burgers\right)}}:\left\{v_1,v_2\right\}_t\left(\boldsymbol{x}\right)\mapsto \left\{v_1,v_2\right\}_{t+\Delta t} \left(\boldsymbol{x}\right).
\label{eq:5}
\end{equation}
The support domain for the input and output functions are the same with that defined in Eq.~(\ref{eq:1}).

Lamé-Navier equation describes the elastodynamics of solids and is widely used in geophysical computing \cite{Liu2017}. The model equation considered here uses the following assumptions. 1) The plain strain assumption, which assumes the deformation identical in the third direction. 2) The deformation is relatively small compared to the range of the considered solid. 3) Elasticity of the considered solid is linear. Then, the Lamé-Navier equation describes the transient deformation of solid as
\begin{equation}
\rho\ddot{\boldsymbol{u}}=\left(\lambda+\mu\right)\nabla\left(\nabla\cdot\boldsymbol{u}\right)+\mu\nabla^2\boldsymbol{u}.
\label{eq:6}
\end{equation}
$\boldsymbol{u}\left(\boldsymbol{x},t\right)=\left\{u_1,u_2\right\}\left(\boldsymbol{x},t\right), \boldsymbol{x}\in\Omega\subset\mathbb{R}^2,t>0,$ is the displacement to be solved, $\lambda$, $\mu$ are Lamé coefficients, $\rho$ is the density of the solid. The time-marching operator of Lamé-Navier equation for LNO to learn is
\begin{equation}
\mathcal{G}_{\rm{L}}^{\rm{\left(L-N\right)}}:\left\{u_1,u_2,{\dot{u}}_1,{\dot{u}}_2\right\}_t\left(\boldsymbol{x}\right)\mapsto \left\{u_1,u_2,{\dot{u}}_1,{\dot{u}}_2\right\}_{t+\Delta t}\left(\boldsymbol{x}\right).
\label{eq:7}
\end{equation}
The support domains for the input and output functions are the same with that defined in Eq.~(\ref{eq:1}) as well. Note that for second-order transient system, the time derivatives ${\dot{u}}_i$ are also unknows to be solved, while their initial values are also required to determine a solution. \ref{secA1} gives an understanding of LNO modelling for the Lamé-Navier equation that, it learns integration operator of Green's function, which is proved to be local-related and shift-invariant.

\subsection{Baseline training and validation\label{sec2:3}}

We train and validate the LNO models separately in learning the two model equations for studying the generalizability.
Parameters, data generation, and training settings for the model equations are as follows. 
For 2D viscous Burgers equation, we set viscosity $\nu=0.01$.
Demo problems for data generation use random initial conditions.
The initial velocities in two directions $\left\{v_1,v_2\right\}$ are individually generated following Eq.~(\ref{eq:3}). 
These demo problems are solved by linear FEM with implicit Euler time-marching scheme. 
For Lamé-Navier equation, we refer to geophysical parameters \cite{Liu2017} that $\lambda=2.71\times{10}^9~\rm{Pa},\,\mu=2.65\times{10}^9~\rm{Pa},\,\rho=2\times{10}^3~\rm{kg/m^3}$. 
The initial displacement in two directions $\left\{u_1,u_2\right\}_{t=0}$ are individually generated as well, while the initial velocities $\left\{{\dot{u}}_1,{\dot{u}}_2\right\}_{t=0}$ are set to 0. 
These demo problems are solved by linear FEM with implicit Newmark $\beta$ time-marching method. 
For each of the two model equations, we generate field series starting from 225 initial conditions as data samples. 
200 of them are used for training and the rest 25 are reserved for validation. 

\begin{figure}[htp]
\centering
\vspace{-70pt}
\includegraphics[width=0.98\textwidth]{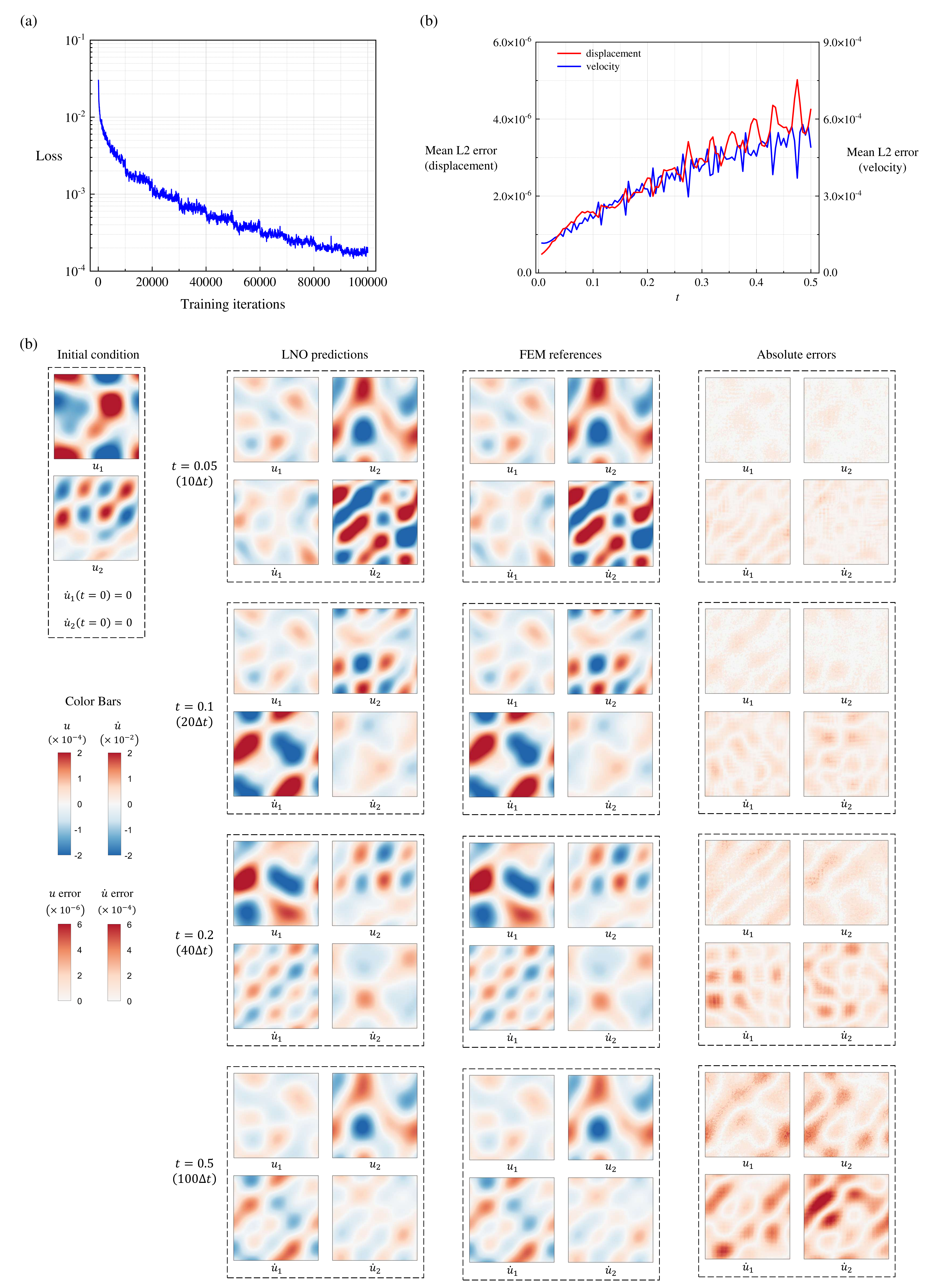}
\caption{The baseline LNO training process and validation results. 
(a) The loss decreases and converges during training. 
(b) The error curve averaged on validation samples. $0.5=100\Delta t$. 
(c) The demo fields of one validation sample, including the randomly generated initial condition, groups of four fields predicted by LNO, the FEM reference (ground truth), and the absolute errors at four time levels.}
\label{fig:baseline_results}
\end{figure}

The two models are both trained following a recurrent supervised schedule \cite{LiHongyu2022}. The Adam optimizer \cite{Kingma2015} is used, while the initial learning rate is set as 0.001 and is manually multiplied by 0.7 every 10,000 iterations. The baseline training results for 2D viscous Burgers equation refer to~\cite{LiHongyu2022}, while the training profile and demo validation fields for Lamé-Navier equation are shown in Fig.~\ref{fig:baseline_results}. 
We quantify the solving error by integrating the averaged error until time $T$ on domain $\Omega$. 
For uniform discretization, the integrated mean error can be approximately calculated as summations that
\begin{equation}
\begin{split}
\mathcal{E}&=\frac{1}{TS_\Omega}\int_0^T\iint_\Omega\lVert \boldsymbol{u}_t\left(\boldsymbol{\xi}\right)-\boldsymbol{\tilde u}_t\left(\boldsymbol{\xi}\right)\lVert_2\mathrm{d}\boldsymbol{\xi}\mathrm{d}t\approx\frac{1}{N_tN_s}\sum_{\tau=1}^{N_t}\sum_{i=1}^{N_s}\lVert \boldsymbol{u}_\tau\left(\boldsymbol{\xi}_i\right)-\boldsymbol{\tilde u}_\tau\left(\boldsymbol{\xi}_i\right)\lVert_2,
\label{eq:8}
\end{split}
\end{equation}
where $S_\Omega$ is the area of $\Omega$, $N_t$ and $N_s$ are the numbers of discretized points in time and space, respectively.
All the fields here are discretized using Cartesian grid. 
We use $\Delta x=1$ for the Lamé-Navier equation. 
Fields used for baseline training and validation are all on square domain with 128$\times$128 nodes. 
$\boldsymbol{u}_t$ and ${\tilde{\boldsymbol{u}}}_t$ are the real solution and the LNO predicted solution, $\boldsymbol{u}_\tau\left(\boldsymbol{\xi}_i\right)$ and ${\tilde{\boldsymbol{u}}}_\tau\left(\boldsymbol{\xi}_i\right)$ are their descretized value at time $\tau$ and position $\boldsymbol{\xi}_i$, respectively.
Errors of different physical fields are discussed separately, as they could have different scales and mean differently. 
For the present validation, the averaged error in learning Lamé-Navier equation are: $5.03\times{10}^{-4}$ (0.511\% of the maximum) for displacement and $7.06\times{10}^{-2}$ (0.489\% of the maximum) for velocity. 
The errors are relatively small. After that, we obtain pre-trained LNO models (denoted as $\mathcal{G}_{\theta^\ast}$) and is ready to study its generalizability.

\section{Evaluating the generalizability of pre-trained LNO}\label{sec3}

One primary difference between classic neural networks and neural operators is the definition space for the input and output. 
Classic neural networks usually define input/output vectors with finite dimensions within bounded value domains, e.g., pixel values for images \cite{Lecun1998}, semantic embedding for languages \cite{Bengio2003}; 
while the input for neural operators uses spaces with infinite dimensions and unbounded value domains in concept. 
This brings difficulties to directly use classic evaluating metrics to the present operator learning task, and the observed general rules could become invalid. 
That is the issue we are facing here.

\subsection{Generalization error parameterized by scalar features}\label{sec3:1}

This study starts from the classic generalization error \cite{Li2012} defined as 
\begin{equation}
\mathcal{E}_{\mathrm{gen}}\left(\mathcal{M}\right)=\mathrm{E}_{\left(X,Y\right)\sim\mathcal{P}}\left[\mathcal{L}\left(Y,\mathcal{M}\left(T\right)\left(X\right)\right)\right],
\label{eq:9}
\end{equation}
where $\mathcal{M}\left(T\right)$ is the statistical learning model trained on dataset $T=\left\{\left(X_i,Y_i\right)\right\}_{i=1}^N$ including $N$ samples.
$\mathcal{L}$ is a scalar-valued error function. 
$X$ and $Y$ are random variables respectively representing the input and output functions. 
$\mathcal{P}$ is a joint distribution for the input-output pairs, while all the training and testing samples are supposed to be i.i.d. from $\mathcal{P}$, i.e., $\left(X_i,Y_i\right)\sim\mathcal{P}$ and $\left(X,Y\right)\sim\mathcal{P}$. 
There appear two mismatches to use this definition for operator learning and specifically the present LNO. 
One is that the \textit{i.i.d.} assumption can hardly be satisfied in practices. 
PDE Problems to be solved for scientific computing are usually discussed by instances, instead of by distributions. 
The other is that, the expectation $\mathrm{E}_{\left(X,Y\right)\sim\mathcal{P}}\left[\cdot\right]$ is hard to estimate for distributions with variable $\left(X,Y\right)$ of infinite dimensions. 

We handle the two issues by making the following modifications to the definition of the generalization error. 
Firstly, we assume that the output function is unique for a given input function, i.e., $Y=\mathcal{G}_\mathrm{L}\left(X\right)$, thus, the joint distribution depends solely on the input function ($X\sim\mathcal{P}$).
Secondly, we use scalar features $\mathcal{F}_X$ for representing the input function $X$, thus reducing the number of input dimensions from infinite to single and making the evaluation of generalization error practical. 
Also, as we discuss facing applications using pre-trained models, statistical features of the training dataset are excluded from the definition for simplification. 
Then, we rewrite the generalization error as
\begin{equation}
\mathcal{E}_{\mathrm{gen}}\left(\mathcal{F}_X;\mathcal{G}_{\theta^\ast}\right)=\mathrm{E}_{X\sim\mathcal{P}\left(\mathcal{F}_X\right)}\left[\mathcal{L}\left(\mathcal{G}_\mathrm{L}\left(X\right),\mathcal{G}_{\theta^\ast}\left(X\right)\right)\right],
\label{eq:10}
\end{equation}
where $\mathcal{E}_{\rm{gen}}\left(\mathcal{F}_X;\mathcal{G}_{\theta^\ast}\right)$ describes the generalization error of pre-trained LNO model $\mathcal{G}_{\theta^\ast}$ with respect to a scalar feature $\mathcal{F}_X$ of function $X$. 
This expectation can be estimated by calculating the averaged error on validation samples with constant $\mathcal{F}_X$ according to Eq.~(\ref{eq:8}). 
There are two main purposes for this definition.
One is to reduce the infinite dimensions of the input to finite or single. 
The other is to quantify the distance between to input functions by the scalar features. 
Thus, we can discuss the generalizability of pre-trained models in view of inter/extrapolation. 

Next, we select representing feature $\mathcal{F}_X$ for Eq.~(\ref{eq:10}). 
Intuitively, amplitude and frequency are primary features of scalar-valued input functions (vector such as the velocity is decomposed in directions as scalars). 
As suggested in \cite{LiHongyu2022}, sample fields for LNO training and validation are generated from random initial conditions defined by Eq.~(\ref{eq:3}). 
By introducing scaling factors, Eq.~(\ref{eq:3}) can be modified and represented as 
\begin{equation}
u_0\left(x_1,x_2\right)=a\boldsymbol{B}^T\left(k_1x_1\right)\boldsymbol{\Lambda}\boldsymbol{B}\left(k_2x_2\right),\qquad x_1,x_2\in\left[1,-1\right],\,a,k_1,k_2\in\mathbb{R}^+,
\label{eq:11}
\end{equation}
where $a$ controls the amplitude. $k_1$ and $k_2$ control the frequency components in the two directions, respectively. 
It is more commonly called wavenumber for spatial directions (the number of waves within a unit spatial range). 
Then, we can investigate the generalization errors of pre-trained LNO models in processing inputs with varying amplitude or wavenumber components.

\subsection{Amplitude parameterized generalization errors}\label{sec3:2}

Amplitude is one of the most fundamental features of functions. 
For functions describing fields in physics, it owns clear physical meanings, such as the speed or distance of movements, or the intensity of pressure. They all play essential roles in computational analyses. 
So firstly, we let $\mathcal{F}_X$ in Eq.~(\ref{eq:10}) equal to the amplitude coefficient $a$ in Eq.~(\ref{eq:11}), then investigate the generalization errors of $\mathcal{G}_{\theta^\ast}$ (the pre-trained LNO model in section 2.3) using the metric $\mathcal{E}_{\rm{gen}}$. 
The followings are experiment settings and results for the two model equations.

\vspace{8pt}
\noindent a)~\textit{In learning viscous Burgers equation}

\begin{figure}[ht!]
\centering
\includegraphics[width=0.9\textwidth]{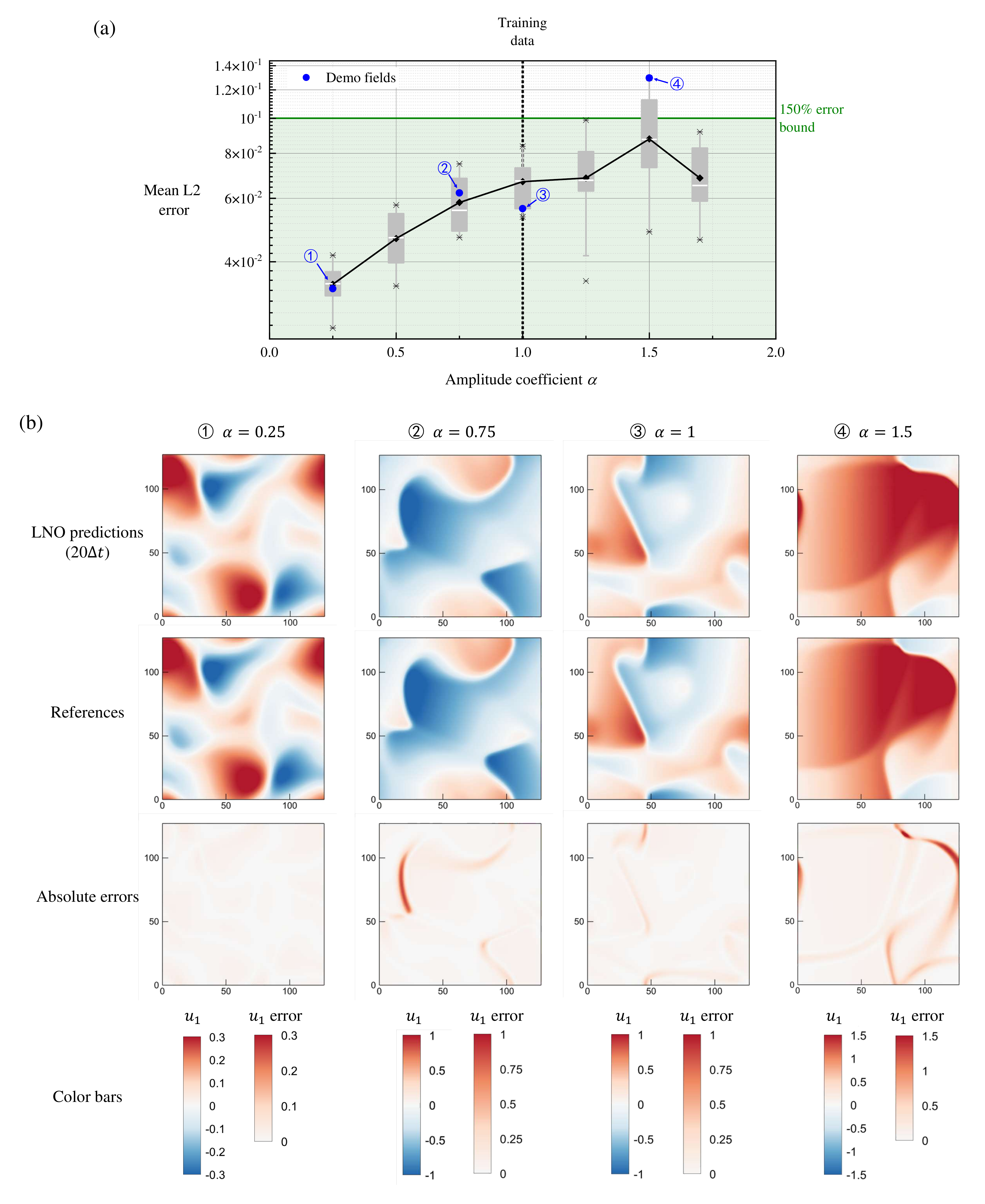}
\caption{The generalization errors of LNO to varied amplitude in learning the viscous Burgers equation. 
(a) The mean L2 error with respect to the amplitude coefficient. 
(b) Demo fields at $t=20\Delta t$ for comparison. These demos are also marked on the curve in (a).}
\label{fig:generalization_results_amplitude_Burgers}
\end{figure}

We let $a=0.25, 0.5, 0.75,1.25,1.5,1.7.$ For each value of $a$, we generate 10 field series from random initial conditions.
Then, combined with the reserved 10 field series ($a=1$), 70 validation field series in total are used for this investigation. 
The generalization errors $\mathcal{E}_{\rm{gen}}$ (normalized by $a$) with respect to function amplitude a are plotted in Fig.~\ref{fig:generalization_results_amplitude_Burgers}(a). 
It can be found that the error grows as the function amplitude become larger.
Within the parameter range of testing, only a relatively small amount of cases exceed the 150\% error bound from the training data, which means the generalizability of pre-trained LNO in amplitude is fine. 
Several example fields are also presented in Fig.~\ref{fig:generalization_results_amplitude_Burgers}(b) to show how physical fields with larger amplitude generate more error. 
For physical systems following the viscous Burgers equation, larger velocity can generate sharper edges that have large gradients, which is hard to predict and is the main error source. 
According to these results, an empirical safe range can be obtained with respect to the amplitude that $\left[a^-,a^+\right]=\left[0.25,1.25\right]$. 
For $a^-$, we use the minimum tested $a$, while it could continue to go down for smaller $a$). 
For $a^+$, only a few samples at $a=1.5$ are observed exceeing the error bound, for example the field \ding{195} in Fig.~\ref{fig:generalization_results_amplitude_Burgers}(b), concerning the increasing overall trend, we use $a^+=1.25$. 
Within this safe range, the pre-trained LNO can give solutions of viscous Burgers equation while keeping the averaged error below 1\% of the maximum field value.

\vspace{8pt}
\noindent b)~\textit{In learning Lamé-Navier equation}

\begin{figure}[htbp]
\centering
\includegraphics[width=0.85\textwidth]{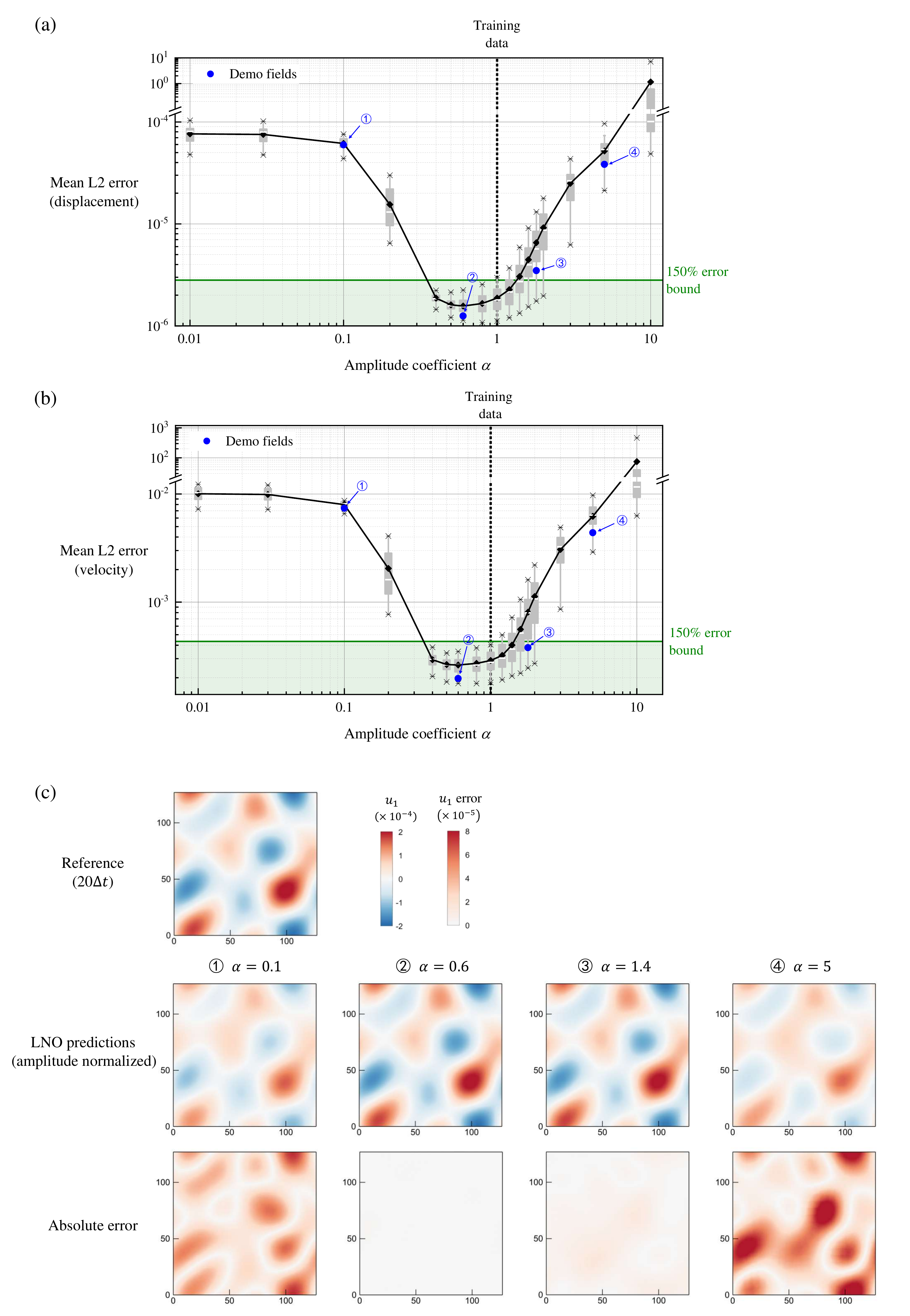}
\caption{The generalization errors of LNO to varied amplitude in learning the Lamé-Navier equation. 
(a) The displacement error. 
(b) The velocity error. 
(c) Demo fields at $t=20\Delta t$ for comparison. 
These demos are also marked on curves in (a) and (b).}
\label{fig:generalization_results_amplitude_LN}
\end{figure}

For homogeneous linear equations such as the present Lamé-Navier equation, if $\boldsymbol{u}\left(\boldsymbol{x},t\right)$ is a solution, then $a\boldsymbol{u}\left(\boldsymbol{x},t\right),\,\\a\in\mathbb{R}$ is a solution as well. 
This feature facilitates the present study on generalizability with respect to the function amplitude $a$, thus we can observe a much wider range of the amplitude coefficiant than that in learning viscous Burgers equation.
Here, we consider 16 different values of $a$ and plot their generalization errors $\mathcal{E}_{\rm{gen}}$ (normalized by $a$) in Fig.~\ref{fig:generalization_results_amplitude_LN}(a) using logarithmic axis. 
It shows a common trend of generalization that, the errors are relatively small around the training data ($a=1$), and they gradually increase as its distance from training data grows. Several example fields are presented in Fig.~\ref{fig:generalization_results_amplitude_LN}(b). 
It shows that whether for the over small or over large function amplitudes, the pre-trained LNO tends to present fake dissipations that make the fields closer to constant fields. 
Notably, we observed a blow-up case when continuously increasing $a$ to 10 (the error rate get over ${10}^5$ times greater for both displacement and velocity), which alerts a strong prohibition to use pre-trained LNO to solve far-extrapolative cases. 
An empirical safe amplitude range $\left[a^-,a^+\right]=\left[0.4,1.2\right]$ can be obtained according to the 150\% error bound from the training data. 
Within this safe range, the pre-trained LNO can give solutions of Lamé-Navier equation while keeping the averaged error below $2.81\times{10}^{-6}$ for displacement (0.56\% of the maximum value) and $4.33\times{10}^{-4}$ for velocity (0.61\% of the maximum value).

\subsection{Wavenumber parameterized generalization errors}\label{sec3:3}

Besides the amplitude, another fundamental feature of functions is ‘shape’, which can be described and parameterized in diverse ways. 
Spectral decomposition is a classic way to describe functions by their shape components, where the higher-order components reflect more intense fluctuations of the functions. 

Similar to that parameterized by amplitudes, we generate additional validation samples using varied wavenumber coefficients. 
Concerning the present data generation from random initial condition according to Eq.~(\ref{eq:11}), we consider the wavenumber coefficients $k_1$ and $k_2$ as representing features of the input function. 
These two varied wavenumber coefficients can be obtained by changing the size of the periodic domain in Eq.~(\ref{eq:3}). 
Defining domain scaling factors in two dimensions as $\left\{\alpha_x,\alpha_y\right\}=\left\{{1}/{k_1},{1}/{k_2}\right\}$, we consider 5 bigger and 5 smaller cases as listed in Table~\ref{tab:varying_wavenumbers}. 
For simplification, we use the mean of the two directions for parameterizing the generalization error $\mathcal{E}_{\rm{gen}}$ as $\mathcal{F}_X=k=\left({k_1+k_2}\right)/{2}$. 
We retain fraction-form scaling factors in the table which show the number of discretized nodes after domain scaling (the original domain is with 128 nodes in each dimension). 
The followings discuss the generalization errors of pre-trained LNO $\mathcal{G}_{\theta^\ast}$ for the two model equations using $\mathcal{E}_{\rm{gen}}$ in Eq.~(\ref{eq:10}) parameterized by the wavenumber coefficient $k$.

\begin{table}[H]
\caption{Varying wavenumbers of the fields by varying sizes of periodic domains.\vspace{-5pt}}
\label{tab:varying_wavenumbers}
\renewcommand\arraystretch{1.5}
\small
\centering
\begin{tabular*}{\textwidth}{@{\extracolsep{\fill}}cccccc@{\extracolsep{\fill}}}
\toprule
\multicolumn{3}{c}{\makecell{\textbf{Group A:} \\ Expand the periodic domain and lower the wavenumber}} & \multicolumn{3}{c}{\makecell{\textbf{Group B:} \\Shrink the periodic domain and higher the wavenumber}} \\
\cmidrule{1-3} \cmidrule{4-6}
\multicolumn{2}{c}{Scaling factor} & \multirow{2}{*}{\makecell{mean wavenumber $k$}} & \multicolumn{2}{c}{Scaling factor} & \multirow{2}{*}{\makecell{mean wavenumber $k$}} \\
\cmidrule{1-2} \cmidrule{4-5}
fractional & decimal & & fractional & decimal	& \\
\midrule
$\displaystyle\left(\frac{220}{128},\frac{220}{128}\right)$ & $\left(1.719,1.719\right)$	& 0.582	& $\displaystyle\left(\frac{90}{128},\frac{128}{128}\right)$ & $\left(0.703,1\right)$	& 1.211 \\
$\displaystyle\left(\frac{180}{128},\frac{220}{128}\right)$	&$\left(1.406,1.719\right)$&	0.646 &		$\displaystyle\left(\frac{72}{128},\frac{128}{128}\right)$&	$\left(0.563,1\right)$	&1.389 \\
$\displaystyle\left(\frac{180}{128},\frac{180}{128}\right)$	&$\left(1.406,1.406\right)$&	0.711 &		$\displaystyle\left(\frac{90}{128},\frac{90}{128}\right)$&	$\left(0.703,0.703\right)$	&1.422\\
$\displaystyle\left(\frac{128}{128},\frac{220}{128}\right)$	&$\left(1,1.719\right)$&	0.791 &		$\displaystyle\left(\frac{72}{128},\frac{90}{128}\right)$&	$\left(0.563,0.703\right)$	&1.6\\
$\displaystyle\left(\frac{128}{128},\frac{180}{128}\right)$	&$\left(1,1.406\right)$&	0.856 &	$\displaystyle\left(\frac{72}{128},\frac{72}{128}\right)$&	$\left(0.563,0.563\right)$	&1.778 \\
\bottomrule
\end{tabular*}
\end{table}

\vspace{8pt}
\noindent a)~\textit{In learning viscous Burgers equation}

\begin{figure}[ht!]
\centering
\includegraphics[width=0.85\textwidth]{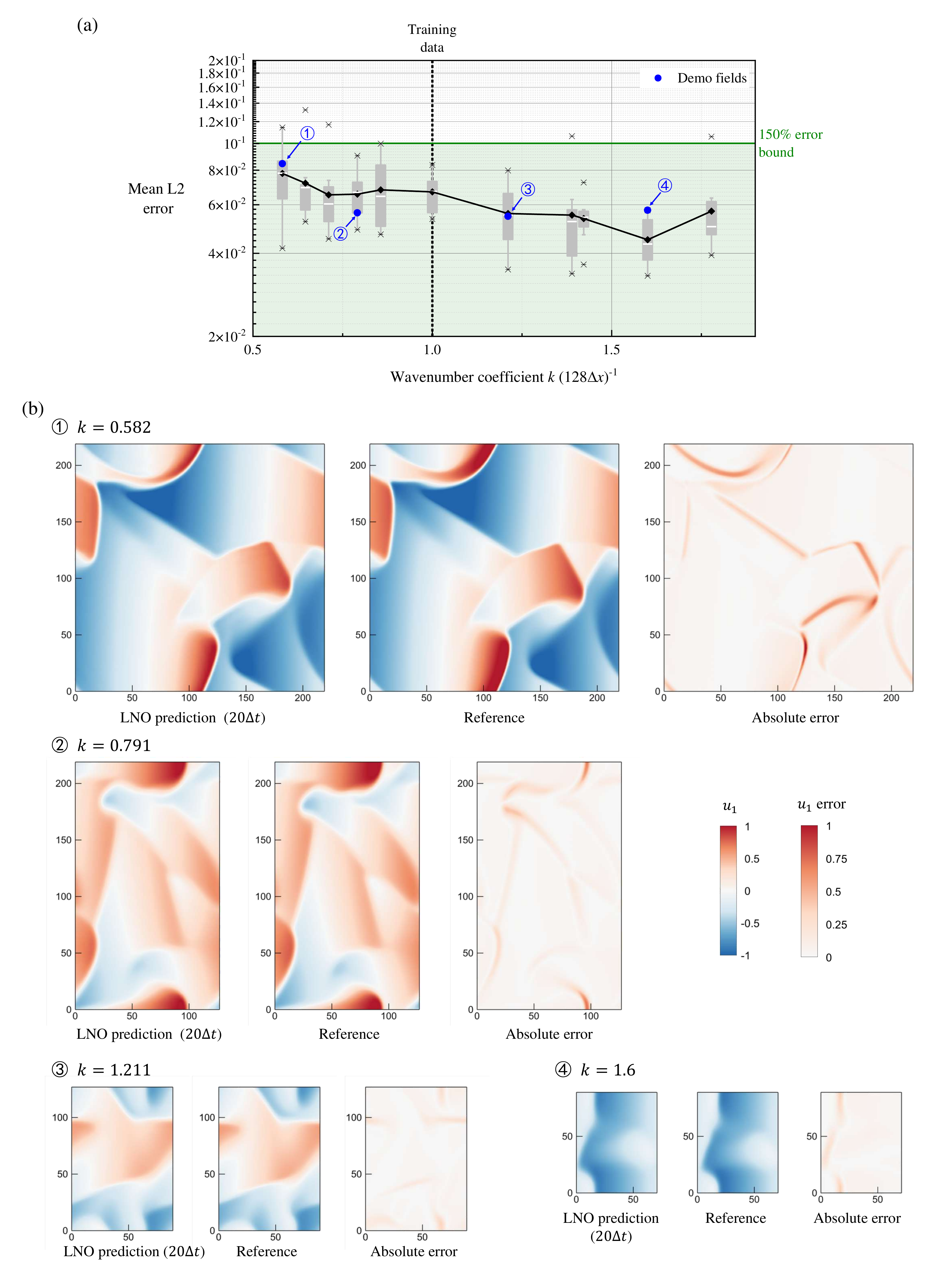}
\caption{The generalization errors of LNO to varied wavenumbers in learning the viscous Burgers equation. 
(a) The mean L2 error with respect to the wavenumber coefficient. 
(b) Demo fields at $t=20\Delta t$ for comparison. These demos are also marked on the curve in (a).}
\label{fig:generalization_results_wavenumber_Burgers}
\end{figure}

We additionally generate viscous Burgers field series starting from 10 extra initial conditions for the ten cases listed in Table~\ref{tab:varying_wavenumbers}. 
Combined with the reserved 10 field series ($k=k_1=k_2=1$), 110 validation field series in total are used to test the pre-trained LNO. 
The generalization errors $\mathcal{E}_\mathrm{gen}$ with respect to function wavenumber $k$ in learning viscous Burgers equation are plotted in Fig.~\ref{fig:generalization_results_wavenumber_Burgers}(a). 
The results show almost a constant error within the considered wavenumber scope form 0.582 to 1.778. 
The range of the mean errors biasing from the training case is within 50\% (maximum: $7.79\times{10}^{-2}$, minimum: $4.48\times{10}^{-2}$), which is very small and nearlly all cases are within the control of 150\% error bound from the training data. 
This implies a quite different generalization rule that the pre-trained LNO seems can be well generalized to predict field series that start from initial condition with a wide range of representing wavenumber. 
The given demo fields in Fig.~\ref{fig:generalization_results_wavenumber_Burgers}(b) further confirmed this observation. 
It can be seen that whether the domain size is, the prediction error of $\mathcal{G}_{\theta^\ast}$ stays at a low level.

\vspace{8pt}
\noindent b)~\textit{In learning Lamé-Navier equation}	

\begin{figure}[htbp]
\centering
\vspace{-70pt}
\includegraphics[width=0.85\textwidth]{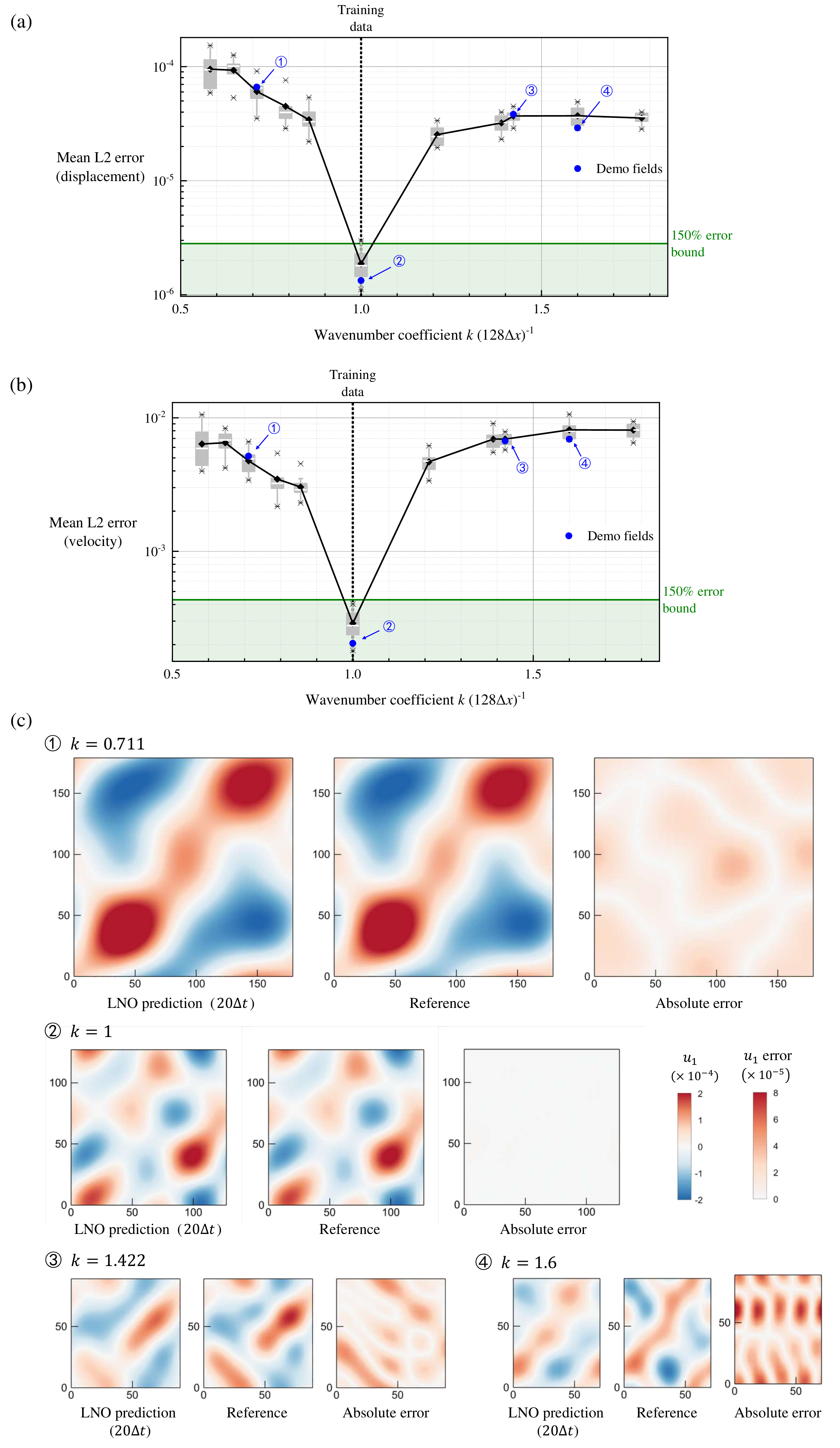}
\caption{The generalization errors of LNO to varied wavenumbers in learning the Lamé-Navier equation. 
(a) The displacement error. 
(b) The velocity error. 
(c) Demo fields at $t=20\Delta t$ for comparison. These demos are also marked on curves in (a) and (b).}
\label{fig:generalization_results_wavenumber_LN}
\end{figure}

We similarly generate validation samples for learning Lamé-Navier equation. 
The same ten cases listed in Table~\ref{tab:varying_wavenumbers} with varied function wavenumber $k$ is considered. 
Lamé-Navier field series starting from 10 extra initial conditions are generated for each of the cases. 
These samples are combined with the reserved 10 field series ($k=k_1=k_2=1$), then used to test the pre-trained LNO. 
The generalization errors $\mathcal{E}_\mathrm{gen}$ with respect to function wavenumber $k$ in learning Lamé-Navier equation are plotted in Fig.~\ref{fig:generalization_results_wavenumber_LN}(a). 
Results here are distinctively different compared to the above three. 
When the wavenumber $k$ of the validation samples slightly biased from the training one from 1 to 0.856 or 1.211, the prediction error of displacement grows immediately from $1.87\times{10}^{-6}$ (0.37\%) for about 1600\% and arrives at about $3\times{10}^{-5}$ (5.96\%). 
Several demo fields in Fig.~\ref{fig:generalization_results_wavenumber_LN}(b) further confirmed this observation. 
These results mean in learning Lamé-Navier equation, the LNO model $\mathcal{G}_{\theta^\ast}$ trained on samples generated from initial condition of one wavenumber component can nearly impossible to generalize to another. 

\vspace{8pt}
In summary, the four investigations on the generalizability of pre-trained LNO show three result types. 
The first, which is the most intuitive rule, is observed in the two experiments using amplitude coefficient $a$ as the parameter $\mathcal{F}_X$ for generalization error $\mathcal{E}_\mathrm{gen}$. 
The pre-trained LNO model performs best when using validation samples generated at the same amplitude as the training data. 
The pre-trained model can generalize to different amplitudes to a certain extent.
The prediction error grows gradually while the amplitude of validation samples is getting biased from the training one.

Anomalies are observed when investigating the generalizability using the wavenumber coefficient $k$. 
The generalizing rules go in opposite directions when learning different equations. 
For the viscous Burgers equation, the pre-trained LNO can accurately predict fields starting from diverse wavenumber components. 
However, for Lamé-Navier equations, once the wavenumber components of the initial condition change, the prediction error rises to a very high level, which indicates wrong solutions. 
These anomalies lead us to think more about the characters of the equation to be learned, the sample organization for training and validation, and their relation, thus explaining the observed generalization errors and giving solutions to reduce them.

\section{Improving the generalizability of LNO for elastodynamic equations }\label{sec4}

This section analyzes the poor generalizability of LNO to different wavenumbers and tries to improve it by direct data supplement and model fine-tuning. We note that the opposite generalization rules of LNO in learning the two equations could be due to the inconsistency between data generation and data sampling for training. 
In section 3, the generalization error is parameterized by features of the initial condition according to Eq.~(\ref{eq:11}). 
However, the field series used for LNO training can be sampled starting from random $t$ besides the initial condition $t=0$ \cite{LiHongyu2022}.
Even if we restrict only using the initial condition as the first input to LNO, it has to process input fields of the continuous ten time levels. 

This sampling strategy makes full use of the generated field series and enriches the inputs to LNO. 
However, it also makes the statistical features of the training set not only related to the random initial condition, but also depends on the equation, which governs the evolution of the physical fields. 
In view of operator learning, the equation governs the evolution of function features, then, it affects the statistical features of the training set for LNO. 
Taking these analyses into consideration, we can guess a reason for the opposite generalization rules (discovered in section 3.3) in learning viscous Burgers equation and Lamé-Navier equation that, \textit{the viscous Burgers equation can automatically evolve the wavenumber components rich while the Lamé-Navier equation can not.}

\subsection{Isolated solution modes for elastodynamic equation}\label{sec4:1}

Now, we present analytical evidence for the difference between the two considered equations. 
It is commonly known that elastodynamic equations are substantially wave equations. 
By applying Helmholtz decomposition to the displacement field, $\boldsymbol{u}=\nabla\phi+\nabla\times\boldsymbol{\psi}$, Lamé-Navier equation can be decomposed to two wave equations for scalar and vector potential function $\phi\left(\boldsymbol{x},t\right)$ and $\boldsymbol{\psi}\left(\boldsymbol{x},t\right), \,\boldsymbol{x}\in\mathbb{R}^3,\,t>0$, as 
\begin{equation}
\left\{
\begin{array}{c}
\rho\ddot{\phi}-\left(\lambda+2\mu\right)\nabla^2\phi=0 \\ \rho\ddot{\boldsymbol{\psi}}-\mu\nabla^2\boldsymbol{\psi}=0
\end{array}
\right..
\label{eq:12}
\end{equation} 
Considering the equation for scalar potential for example, and with initial condition, $\phi\left(\boldsymbol{x},t=0\right)=\phi_0$, $\dot{\phi}\left(\boldsymbol{x},t=0\right)=\phi_1$, Kirchhoff’s formula \cite{Evans2010} gives its solution as
\begin{equation}
\phi\left(\boldsymbol{x},t\right)=\oint_{\partial B\left(\boldsymbol{x},t\right)}\left[{\phi_0\left(\boldsymbol{y}\right)+\nabla\phi_0\left(\boldsymbol{y}\right)\cdot\left(\boldsymbol{y}-\boldsymbol{x}\right)+t\phi_1\left(\boldsymbol{y}\right)}\right]\mathrm{d}S\left(\boldsymbol{y}\right).
\label{eq:13}
\end{equation} 
The coefficients are neglected for this qualitative analysis.
$B\left(\boldsymbol{x},t\right)$ is a ball of radius $t$ centered at $\boldsymbol{x}$. 
Considering its Fourier transform and by coordinate substitution $\boldsymbol{y}=\boldsymbol{x}+t\boldsymbol{z}$, we have 
\begin{equation}
\hat{\phi}\left(\boldsymbol{\xi},t\right)=\oint_{\partial B\left(\mathbf{0},1\right)}\left[{\underbrace{\int_{\mathbb{R}^3}{\phi_0\left(\boldsymbol{x}+t\boldsymbol{z}\right)e^{-i\pi\boldsymbol{\xi}\cdot\boldsymbol{x}}d\boldsymbol{x}}}_{{\hat{\phi}}_0\left(\boldsymbol{\xi}\right)e^{2\pi it\boldsymbol{z}\cdot\boldsymbol{\xi}}}}+t\boldsymbol{z}\cdot{\underbrace{\int_{\mathbb{R}^3}{\nabla\phi_0\left(\boldsymbol{x}+t\boldsymbol{z}\right)e^{-i\pi\boldsymbol{\xi}\cdot\boldsymbol{x}}d\boldsymbol{x}}}_{2\pi i\boldsymbol{\xi}{\hat{\phi}}_0\left(\boldsymbol{\xi}\right)e^{2\pi it\boldsymbol{z}\cdot\boldsymbol{\xi}}}}+t{\underbrace{\int_{\mathbb{R}^3}{\phi_1\left(\boldsymbol{x}+t\boldsymbol{z}\right)e^{-i\pi\boldsymbol{\xi}\cdot\boldsymbol{x}}d\boldsymbol{x}}}_{{\hat{\phi}}_1\left(\boldsymbol{\xi}\right)e^{2\pi it\boldsymbol{z}\cdot\boldsymbol{\xi}}}}\right]\mathrm{d}S\left(\boldsymbol{z}\right) 
\label{eq:14}
\end{equation}
It shows that, once ${\hat{\phi}}_0\left(\boldsymbol{\xi}\right)=0$ and ${\hat{\phi}}_1\left(\boldsymbol{\xi}\right)=0$, we have $\hat{\phi}\left(\boldsymbol{\xi},t\right)=0$. 
In other words, Fourier modes in the initial condition do not transfer to each other when evolving following the wave equation. 
If some Fourier components are zero in the initial condition, they will always be zero in the solution. 
We call these Fourier components \textit{isolated} to each other. 
This inference can be easily generalized to Lamé-Navier equation, as divergence and curl calculations will not transfer Fourier components to each other. 
Also, it also stands for 2D cases according to dimension descent method.

\begin{figure}[ht!]
\centering
\includegraphics[width=0.95\textwidth]{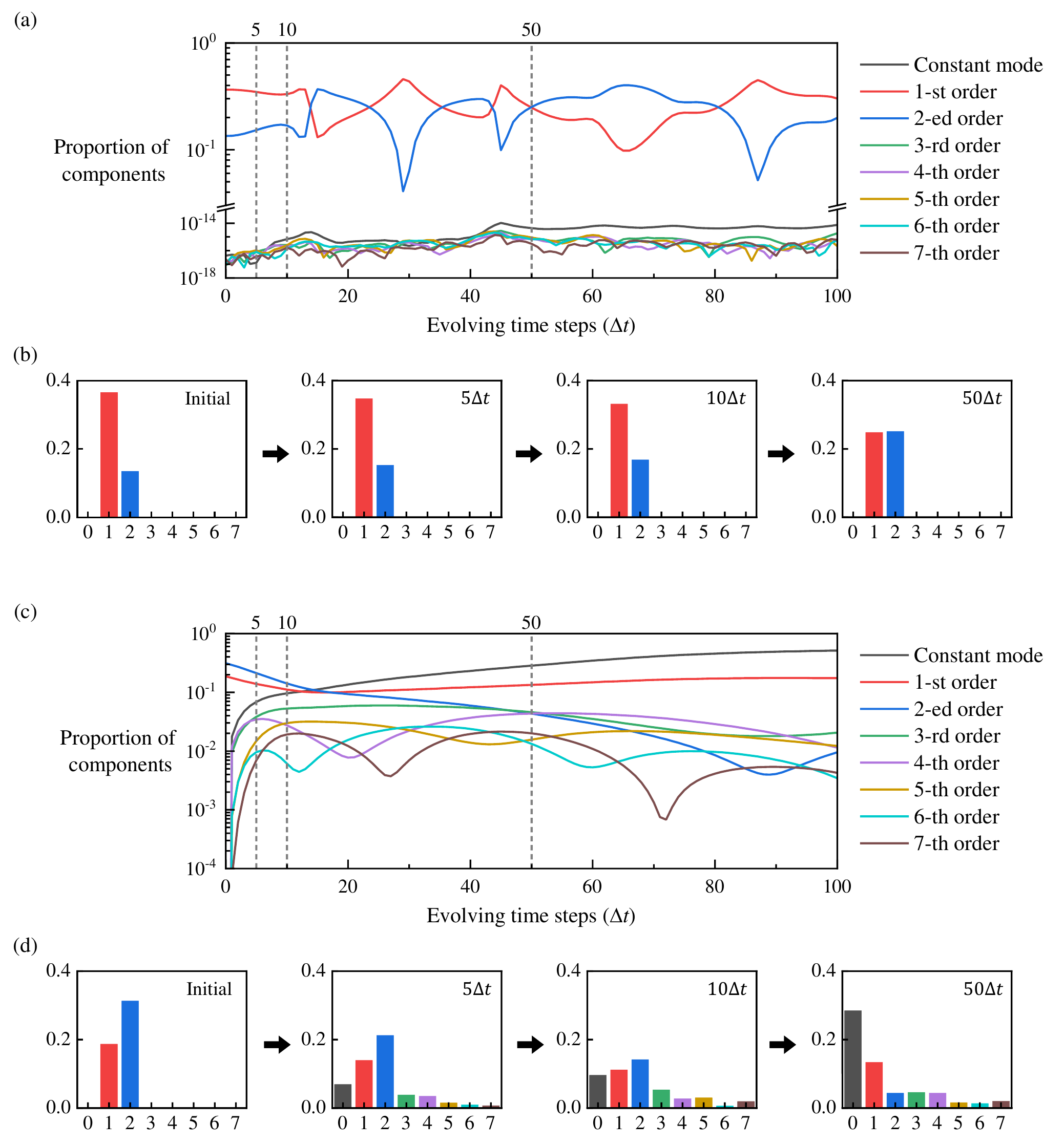}
\caption{Different wavenumber modes evolving rules of different transient equations. 
(a) The wavenumber evolving of Lamé-Navier equation, with (b), its components at four time levels. 
(c) The wavenumber evolving of Burgers equation, with (d), its components at four time levels plotted.}
\label{fig:mode_evolving}
\end{figure}

For intuitive demonstration, we take one field series from Lamé-Navier equation dataset and monitor the evolution of its Fourier components. 
We conduct fast Fourier transform in one direction and integrate the components in the other for simplification. 
The evolving trajectories of the first eight components are shown in Fig.~\ref{fig:mode_evolving}(a). 
It is seen that the initial condition includes the first and the second order components as it is randomly generated according to Eq.~(\ref{eq:11}). 
The evolving trajectories are consistent to the above analyses that other components hold on zero throughout the evolving process. This means that the Fourier modes do not transfer to others. 
Fig.~\ref{fig:mode_evolving}(b) shows the histograms at $t=0,\, 5\Delta t,\,10\Delta t,\,50\Delta t$. It indicates that amplitudes of other modes are all within the machine-error range below ${10}^{-14}$. 

For comparison, the same analyses are conducted on a field series from viscous Burgers equation dataset, the results are in Figs.~\ref{fig:mode_evolving}(c) and \ref{fig:mode_evolving}(d).
Distinctly different evolving trajectories are shown that, the first and the second order components transfer to others. 
As evolving time goes forward, the higher-order components gradually diffused to the zeroth order owing to the viscosity. 
The transfer of Fourier components should be attribute to the nonlinear item $\boldsymbol{v}\cdot\nabla\boldsymbol{v}$. 
Otherwise, if we solely consider linear diffusion equation (also known as heat equation) and its initial-value problem
\begin{equation}
\dot{v}\left(\boldsymbol{x},t\right)=\Delta v\left(\boldsymbol{x},t\right),\qquad v\left(\boldsymbol{x},t=0\right)=v_0\left(\boldsymbol{x}\right),\, \boldsymbol{x}\in\mathbb{R}^2,\,t>0.
\label{eq:15}
\end{equation}
The solution and its Fourier transform are \cite{Evans2010,Stein2003}
\begin{equation}
\left(\boldsymbol{x},t\right)=\frac{1}{4\pi t}\iint_{\mathbb{R}^2}{e^{-\frac{\left|\boldsymbol{x}-\boldsymbol{y}\right|^2}{4t}}v_0\left(\boldsymbol{y}\right)\mathrm{d}\boldsymbol{y}},\qquad\hat{v}\left(\boldsymbol{\xi},t\right)={\hat{v}}_0\left(\boldsymbol{\xi}\right)e^{-4\pi^2\left|\boldsymbol{\xi}\right|^2t}.
\label{eq:16}
\end{equation}
It shows that, if ${\hat{v}}_0\left(\boldsymbol{\xi}\right)=0$, then $\hat{v}\left(\boldsymbol{\xi},t\right)=0$, i.e., Fourier components of the solutions for linear diffusion equation are also \textit{isolated} which is similar to that derived from the wave equation. 

In summary, the above analyses and results confirmed the different evolving natures of the two equations. 
The isolated solution modes of Lamé-Navier equation (essentially wave equations) limit wavenumber modes in the dataset to the first and second order according to the random initial condition Eq.~(\ref{eq:11}). 
In contrast, solution modes of viscous Burgers equation evolve automatically via the nonlinear term, thus resulting in rich field modes in datasets and great generalizability of the pre-trained LNO model to different wavenumber modes. 
These inferences suggest that, the generalizability of pre-trained LNO for Lamé-Navier equation can be improved by supplementing data in other wavenumbers.

\subsection{Fine-tuning with supplemented solution modes}\label{sec4:2}

\begin{figure}[htbp]
\centering
\includegraphics[width=0.6\textwidth]{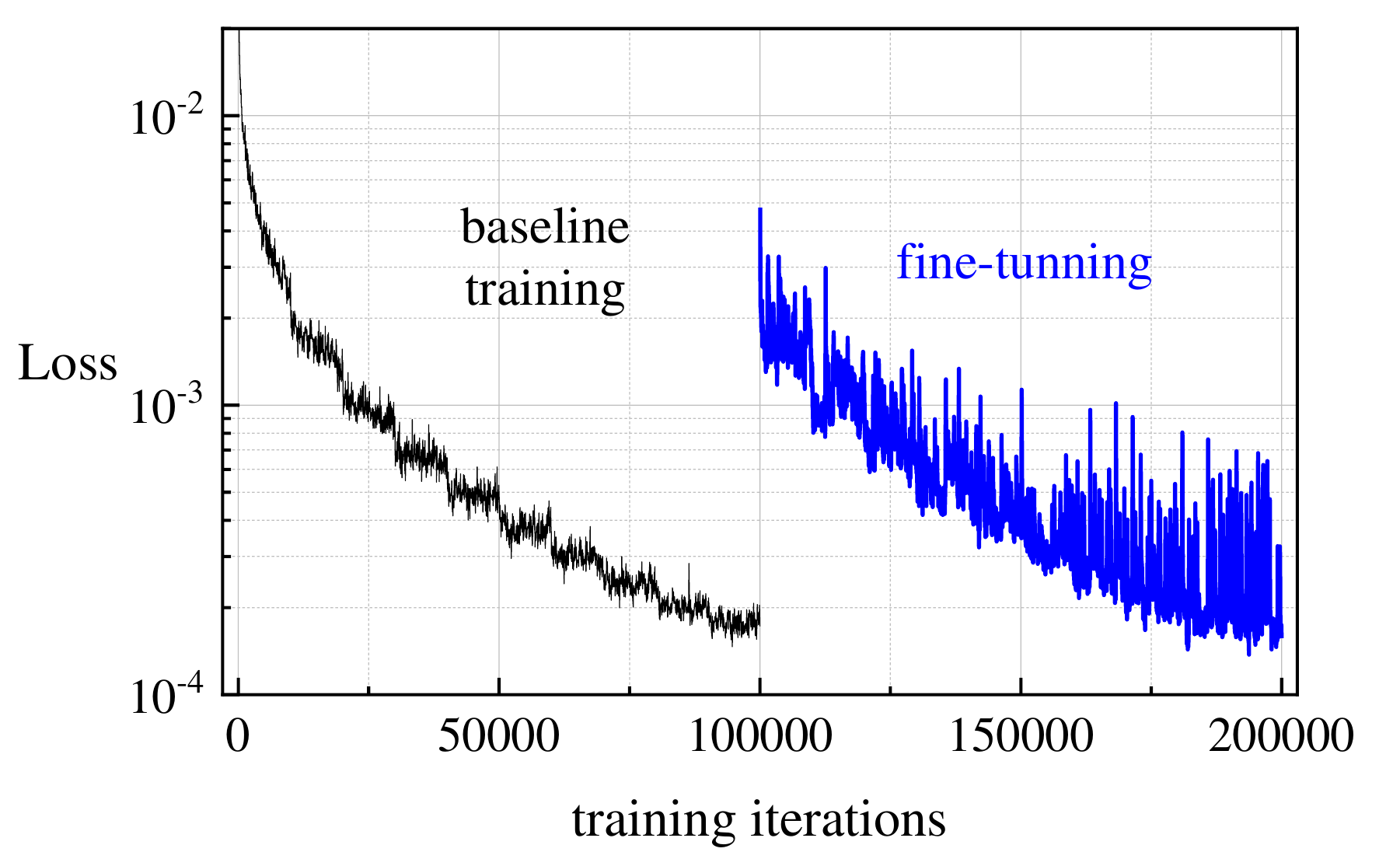}
\caption{The two-stage loss curve of LNO training.}
\label{fig:fine_tunning_loss}
\end{figure}

Though the Lamé-Navier equation can not evolve wavenumber component rich by itself, it can be manually supplemented by improving the data generation strategy. 
Similar to that used for investigating the generalization rule of diverse frequencies, we additionally generate field series from 50 random initial conditions for each of the ten cases listed in Table~\ref{tab:varying_wavenumbers}.
The pre-trained LNO is fine-tuned using these supplemented data. 
The fine-tuning starts from the trained baseline model, while the optimizer and training settings are identical to the baseline training. 
One of the training trajectories is shown in Fig.~\ref{fig:fine_tunning_loss}. 
It shows that, as the fine-tuning uses data in more variety, which is harder for LNO to fit, the training fluctuates harder than the baseline trajectory. 
Still, it converged successfully.

We conduct the fine-tuning experiment with two tests to reveal as much of the generalization and data supplementation behavior of LNO as possible. 
The two fine-tuning tests both start from the trained baseline model but use different supplementary data. 
Test A only uses extended data of smaller wavenumber (the first block in Table~\ref{tab:varying_wavenumbers}). 
Test B uses all the extended data. 
The fine-tuned LNO models are validated using 10 samples for each wavenumber. 
These validation samples are generated from new random initial conditions to keep them unused in training.

\begin{figure}[ht]
\centering
\includegraphics[width=0.8\textwidth]{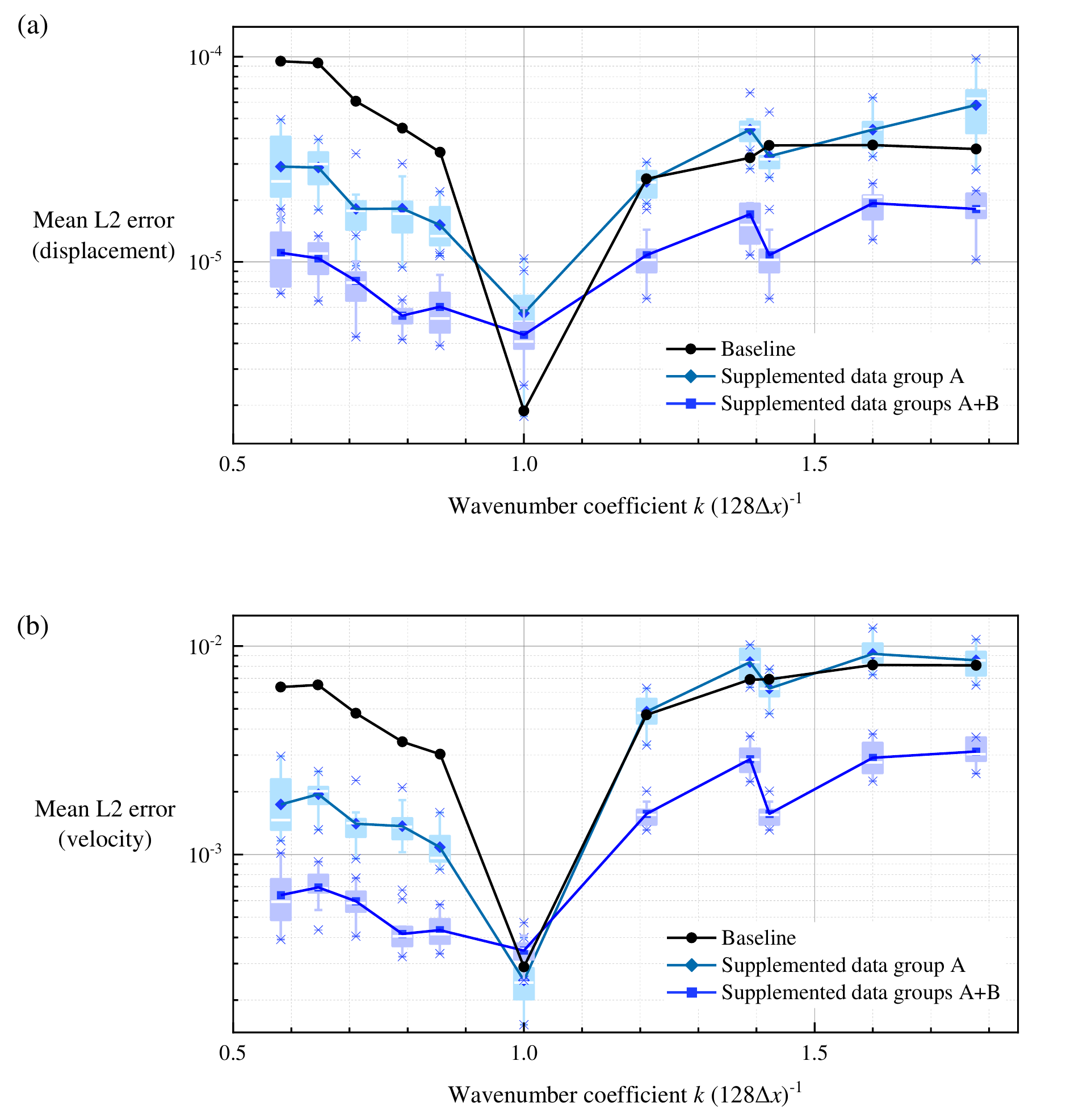}
\caption{The validation error comparison before and after fine-tuning. 
Two tests using different supplemented data groups are compared. 
(a) Displacement errors. 
(b) Velocity errors.}
\label{fig:fine_tunning_comparison}
\end{figure}

Validation errors of the two tests are plotted in Fig.~\ref{fig:fine_tunning_comparison} where subfigure (a) for the displacement and subfigure (b) for the velocity. 
These results display obvious improvements in reducing the generalization error that, the errors when $k<1$ are reduced by the LNO fine-tuning test A, and all the generalization errors are reduced by fine-tuning test B. 
These results are consistent with the usual rule that, the generalizability of pre-trained LNO to diverse wavenumbers can be effectively improved by data supplement and fine-tuning. 
Besides, these results bring out two interesting and worth-discussing phenomena. 
The first is that, though part of the $k=1$ samples (the baseline dataset) are reserved for fine-tuning, the fine-tuned LNO model lost its accuracy a bit on the baseline validation case ($k=1$). 
This implies a limited approximate capability of LNO without supplementary architecture or additional learnable weight. 
If we require one LNO model to learn the dynamics of different wavenumber components, the approximation quality for the dynamics of each single component could be lost. 
The other phenomenon is that, when comparing the error of test B to test A, samples with greater wavenumber for test B not only reduce the error of $k>1$ cases but also help the generalization performance of $k<1$ cases. 
We understand this behavior as that, when the diversity of data is poor to depict a hidden general law, the neural networks tend to \textit{memorize} each of the isolated cases to get lower loss in training; 
as the diversity increases and exceeds a threshold value, the neural networks have to \textit{understand} the hidden law, which is harder but could be able to make the model perform better. 
This phenomenon is similar to the \textit{emergent ability} in large language model research \cite{Wei2022}, but it appears with the increasing variety of data, instead of model size. 
Or this phenomenon can be interpreted as the ‘\textit{grokking}’ of neural networks, referring to the latest advances in deep learning technology \cite{Power2022,Ziming2022}.

\section{Elastodynamic applications}\label{sec5}

Finally, we use pre-trained LNO models for Lamé-Navier equation to solve elastodynamic model problems.
On the one hand, these application examples demonstrate the necessity of improving the generalizability of LNO to different wavenumbers. 
The baseline LNO model trained in section 2.3 and that fine-tuned in section 4.2 are compared here. 
On the other hand, these demo samples broaden the application scope of LNO to elastodynamic application, which models the occurrence of earthquakes and their wave propagations \cite{Aki2002}. 
Conventionally, its numerical solving requires giant computer source \cite{Fu2017} which could be reduced by using the present intelligent computation framework. 
Before presenting the results, we introduce the solving workflow as follows.

\subsection{Solving workflow using pre-trained LNO}\label{sec5:1}

\begin{figure}[ht]
\centering
\includegraphics[width=1\textwidth]{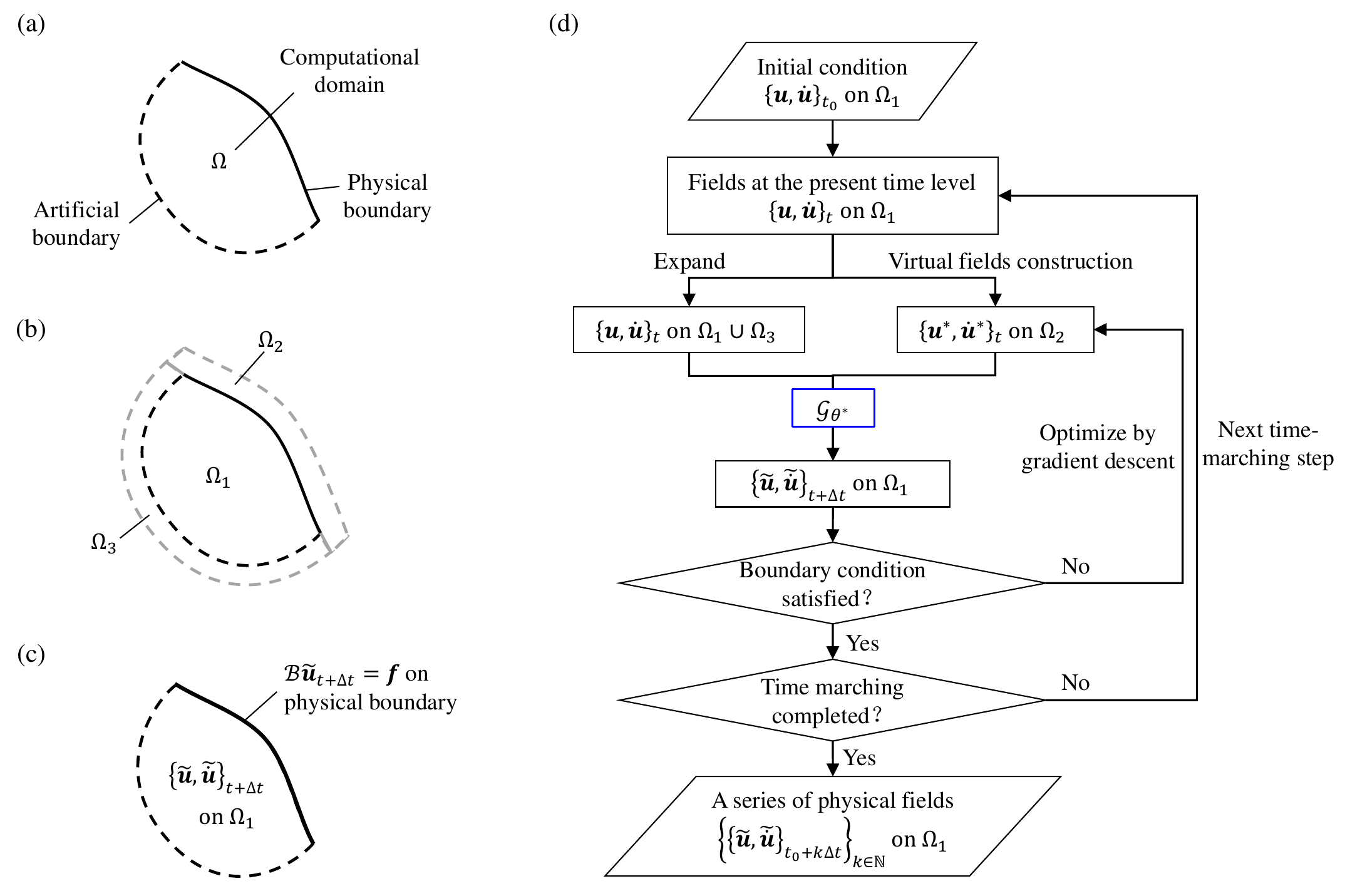}
\caption{The solving workflow using pre-trained LNO $\mathcal{G}_{\theta^\ast}$. 
(a) Domain and boundaries of the considered general problem. 
(b) Domain expansion and division. 
(c) Output fields from $\mathcal{G}_{\theta^\ast}$ and the boundary condition to be satisfied by optimization.
(d) The overall workflow. }
\label{fig:workflow}
\end{figure}

Following the conception of LNO~\cite{LiHongyu2022} and within its reusable scope, the pre-trained LNO model can solve problems in this section without additional training or fine-tuning. 
The solving workflow consists of three parts as shown in Fig.~\ref{fig:workflow}. The first part is domain preparation, the computational domain should be divided into near-boundary and away-boundary areas.
The second part is boundary imposing, the general principle is to operate on the near-boundary areas to make the solution fields satisfy the boundary condition as far as possible. 
The specific imposing techniques can be different depending on problems and boundary types. Examples include outward domain padding, immersed boundary method \cite{LiHongyu2022}, and virtual domain extension (VDE) by solving optimization problems \cite{Ye2025}. 
The third part is solving, the time series solutions are obtained by recurrent time marching starting from the initial condition and using the pre-trained LNO model together with boundary conditions properly imposed. 
In each of the time-marching iterations, the LNO model uses the last time-level field on both the near-boundary and away-boundary areas as the input, then, it outputs the next time-level solution on the away-boundary area, while fields on the near-boundary area are given or supplemented for boundary imposing.

\subsection{Elastic wave propagation in infinite space}\label{sec5:2}

\begin{figure}[h]
\centering
\includegraphics[width=1\textwidth]{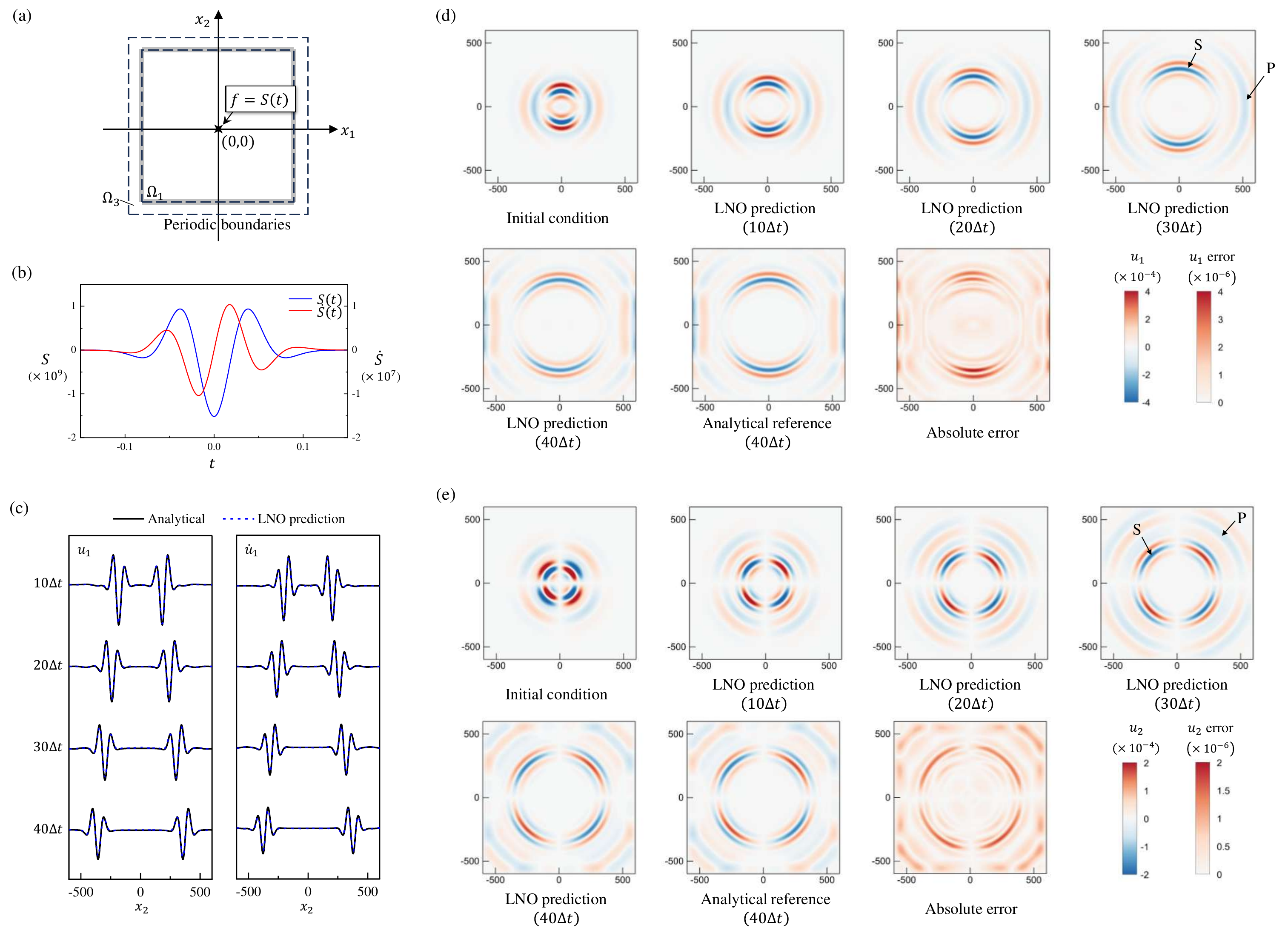}
\caption{The pre-trained and improved LNO model solves the elastic wave propagation in infinite space. 
(a) The model problem and its domain division for LNO solving. 
(b) The transient point source. 
(c) The displacement and velocity profiles on $x_1=0$. 
(d) The predicted $u_1$ fields. 
(e) The predicted $u_2$ fields.}
\label{fig:solve_infinite}
\end{figure}

We first consider the elastic wave propagation in infinite space that is generated by a point source $S(t)$ in direction $i$ and located at the coordinate origin, as shown in as shown in Fig.~\ref{fig:solve_infinite}(a). We use the second time derivative of Ricker’s wavelet as the point source that
\begin{equation}
\begin{split}
S\left(t\right)&=A\left[\left(1-2\pi^2f^2t^2\right)\mathrm{exp}\left({-\pi^2f^2t^2}\right)\right]^{\prime\prime} \\
&=A\left(24\pi^4f^4t^2-8\pi^6f^6t^4-6\pi^2f^2\right)\mathrm{exp}\left({-\pi^2f^2t^2}\right),
\label{eq:17}
\end{split}
\end{equation}
where $f$ decides the frequency of the source. 
Here, we consider three cases of $f=8,12,$ and $18\,\mathrm{Hz}$ to discuss the generalizability to different frequency (wavenumber for spatial dimensions) and to validate our fine-tuning using supplemented data. 
The amplitude coefficient $A$ is set to $\left[6,6,4\right]\times{10}^3$ respectively for the three source frequencies to make the solution amplitude within the safe generalization scope referring to section 3.2. 
For simplification, we use a finite computational domain $\Omega=\left[-600,600\right]^2$ with periodic boundaries as an approximation of the infinite space. 
The analytical solution detailed in \ref{secA2} is used for reference in this case.

Following Fig.~\ref{fig:workflow}, we extend the domain and divide it into $\Omega_1$ (equals the original $\Omega$) and $\Omega_2$ (obtained by periodic extension outward the boundaries).
For each of the time marching step, the pre-trained LNO receives $\boldsymbol{u}_t\left(\boldsymbol{x}\right),\ \boldsymbol{x}\in\Omega_1\cup\Omega_2$ as input, then, output the predicted solution $\tilde{\boldsymbol{u}}_{t+\Delta t}\left(\boldsymbol{x}\right),\boldsymbol{x}\in\Omega_1$. 
This case has no physical boundary, thus there is no $\Omega_3$ and no need to imposing any boundary condition. 
It is noted, LNO learns Lamé-Navier equation with no source term in Eq.~(\ref{eq:6}), yet this model problem is with a transient source. 
For this issue, as the source duration is relatively short, we let LNO solve starting from time $t_0$ that the source end. 
For the Rikers’ wavelet of $f=8$ plotted in Fig.~\ref{fig:solve_infinite}(b), $t_0=0.12$.

We compare the solutions for the case $f=8\, \mathrm{Hz}$. 
The wave propagating process is shown in Figs.~\ref{fig:solve_infinite}(c), \ref{fig:solve_infinite}(d), and \ref{fig:solve_infinite}(e). 
The displacement and velocity fields solved by the fine-tuned LNO are compared to the analytical solution. 
Also, these fields are compared quantitatively by curves on $x_1=0$ in subfigures (c). 
Overall, the results match well. 
Modes of the faster compressional wave (P-wave) and the slower shear wave (S-wave) get apart clearly. 
The main features of this dynamic process such as the shape, amplitude, and position of the wavefront are all correctly captured by the pre-trained LNO. 
In quantity, the averaged absolute displacement and velocity errors on $x_1=0$ are listed in Table~\ref{tab:result_infinite_1}. 
The error rates refer to the corresponding maximum field amplitudes. 
It is seen that the errors are relatively small, under 3\%. 
These results confirmed our LNO modeling and its feasibility to solve elastodynamic problems.

\vspace{10pt}

\begin{table}[H]
\caption{Errors in solving elastic wave propagation in infinite space.\vspace{-5pt}}
\label{tab:result_infinite_1}
\renewcommand\arraystretch{1}
\small
\centering
\begin{tabular*}{\textwidth}{@{\extracolsep{\fill}}cccccc@{\extracolsep{\fill}}}
\toprule
\multicolumn{2}{c}{Time step} & $10\Delta t$ & $20\Delta t$ & $30\Delta t$ & $40\Delta t$ \\
\midrule
\multirow{2}{*}{\makecell{Displacement\\$u_1$}} & Averaged absolute error ($\times{10}^{-7}$) on $x_1=0$
	&2.65	&4.88	&7.17	&7.25\\
\cmidrule{2-6}
& Error rate
(\% of the maximal absolute value)	&0.73	&1.55	&2.50	&2.74\\
\midrule
\multirow{2}{*}{\makecell{Velocity\\$\dot u_1$}} & Averaged absolute error ($\times{10}^{-5}$) on $x_1=0$ 
	&2.09	&2.79	&3.90	&4.65\\
\cmidrule{2-6}
& Error rate
(\% of the maximal absolute value)	&0.72	&1.09	&1.70	&2.21\\
\bottomrule
\end{tabular*}
\end{table}

\begin{figure}[htbp]
\centering
\includegraphics[width=\textwidth]{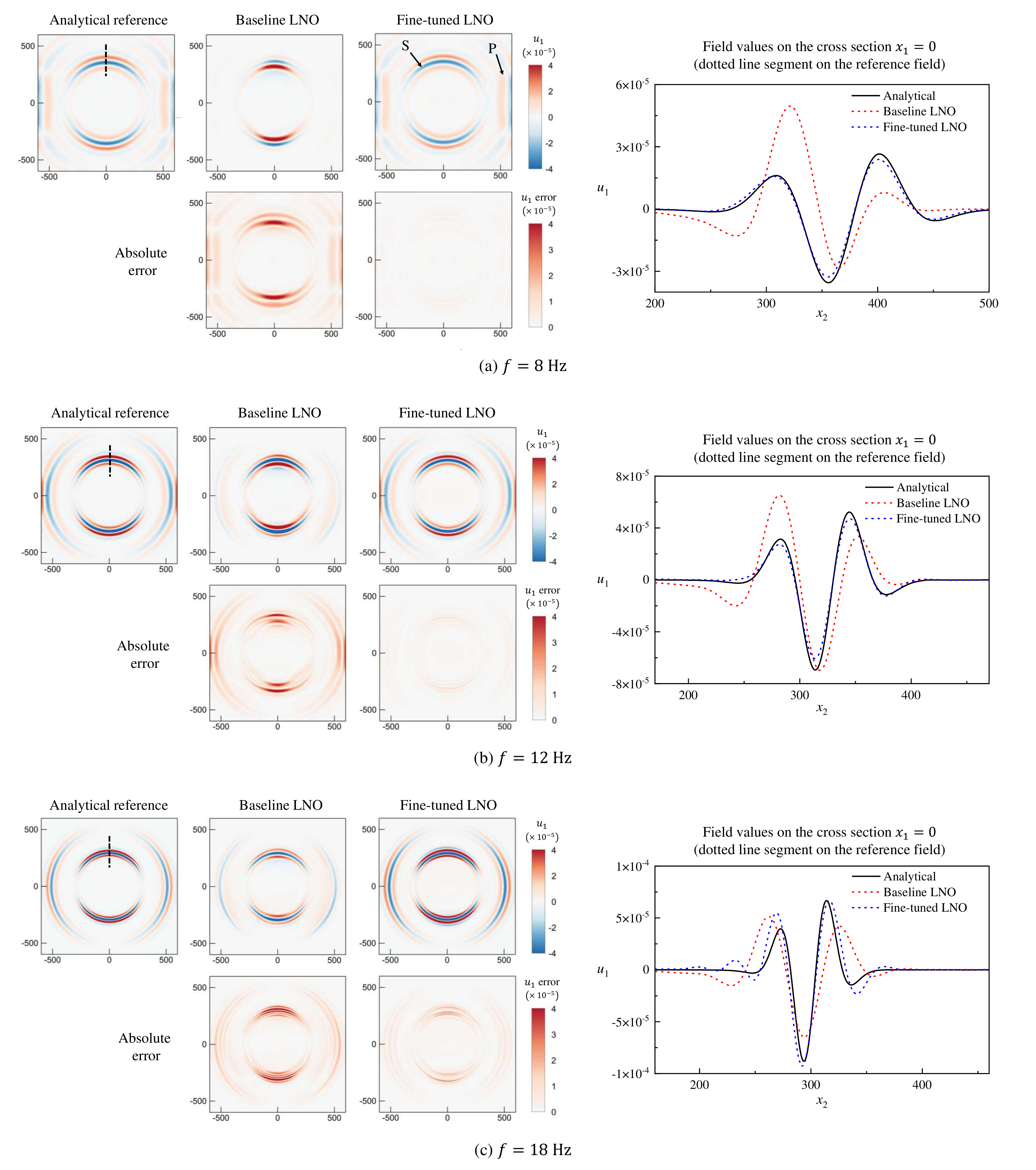}
\caption{Comparisons of the baseline and fine-tuned LNO in solving elastic wave propagations in infinite space excited by sources of three frequencies. 
(a) $f=8\, \mathrm{Hz}$. 
(b) $f=12\, \mathrm{Hz}$. 
(c) $f=18\, \mathrm{Hz}$. 
The fields of displacement $u_1$ are compared on the left, while the field values on the cross section (the dotted lines on the reference fields) are compared on the right.}
\label{fig:solve_infinite_frequencies}
\end{figure}

\begin{figure}[htbp]
\centering
\includegraphics[width=0.8\textwidth]{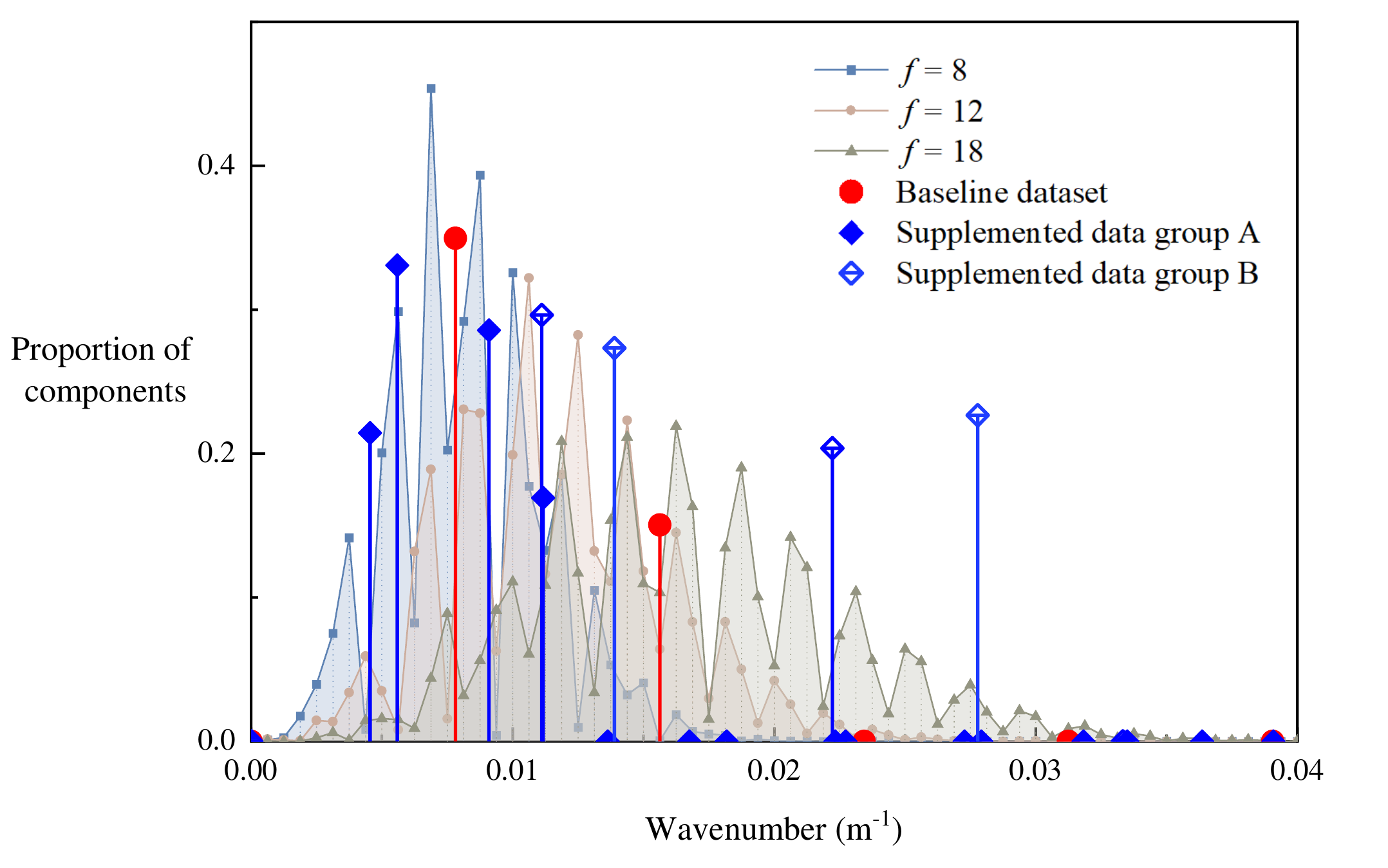}
\vspace{-10pt}
\caption{Relation between wavenumber components of the training data groups and the model problems to be solved (propagation of elastic wave in infinite space excited by sources of three frequencies $f=8,\,12,\,18\,\mathrm{Hz})$.}
\label{fig:frequencies_comparison}
\end{figure}

Then, we compare solutions by LNO before and after data supplement and fine-tuning, as in Fig.~\ref{fig:solve_infinite_frequencies}. 
The analytical solution is still used for reference. 
Solution fields for Ricker frequency $f=8, 12, 18\, \mathrm{Hz}$ after 40 time-marching steps are plotted with typical P-wave and S-wave modes marked. 
Displacement curves on the typical section around the S-wave front (marked on fields by black dotted lines) are plotted and compared on the right. 
It can be seen that, the solutions given by the baseline model are almost wrong. 
The P-wave mode disappeared, and the S-waveform is greatly distorted. 
These errors are all mitigated after data supplement and fine-tuning, by which the pre-trained LNO can solve in a wider range of wavenumbers. 
According to the quantitative results on the wavefront section in Table~\ref{tab:result_infinite_2}, the errors are significantly reduced, with the error of case $f=8\, \mathrm{Hz}$ being reduced the most at 90.8\%. 
To further show the wavenumber relation between the training data and the model problem solved here, we conduct spectral analyses on the displacement fields of the present solution on $x_1=0$ and the training data groups (see Table~\ref{tab:varying_wavenumbers}). 
The decomposed frequencies are plotted in Fig.~\ref{fig:frequencies_comparison}. 
It shows that, the two components in the baseline dataset can hardly cover that of the demo problem. 
By supplementing with the two data groups (smaller wavenumbers and larger wavenumbers), the lack of wavenumber modes is mitigated, thus the solving errors are reduced. 
When looking into the three model problems in comparison, it can be found that the point source with higher frequency generates richer wavenumber components. 
Though the training samples are supplemented, many wavenumber modes of the highest case of $f=18\, \mathrm{Hz}$ are located in the large-wavenumber area that has relatively sparser samples. 
This explains why the fine-tuned LNO can not solve this case as precisely as solving cases of $f=8,12\, \mathrm{Hz}$. 
Even so, it is predictable that more data supplements and fine-tuning can continuously decrease the solving error.

\begin{table}[H]
\caption{The error comparisons before (the baseline) and after (the fine-tuned) data supplementation and fine-tuning in solving elastic wave propagations in infinite space excited by sources of three frequencies.}
\label{tab:result_infinite_2}
\renewcommand\arraystretch{1}
\small
\centering
\begin{tabular*}{\textwidth}{@{\extracolsep{\fill}}ccccc@{\extracolsep{\fill}}}
\toprule
\makecell{Source frequency\\ (Hz)} & & Baseline & Fine-tuned & \makecell{Error reduced \\rate (\%)} \\
\midrule
\multirow{2}{*}{8} & Mean absolute error ~ ($\times{10}^{-6}$)	&11.70	&1.08 & \multirow{2}{*}{90.8} \\
\cmidrule{2-4}
& Error rate (\% of the maximal amplitude) &32.99	&3.04  &\\
\midrule
\multirow{2}{*}{12} & Mean absolute error ~ ($\times{10}^{-6}$)	&10.93	&1.76 & \multirow{2}{*}{83.9} \\
\cmidrule{2-4}
& Error rate (\% of the maximal amplitude) &15.72	&2.53  &\\
\midrule
\multirow{2}{*}{18} & Mean absolute error ~ ($\times{10}^{-6}$)	&11.47	&4.96 & \multirow{2}{*}{56.8} \\
\cmidrule{2-4}
& Error rate (\% of the maximal amplitude) &13.00	&5.62  &\\
\bottomrule
\end{tabular*}
\end{table}

\vspace{10pt}

\vspace{10pt}

\subsection{Lamb problem}\label{sec5:3}
Lamb problem concerns the wave propagation in half-infinite elastomer (the earth media) excited by external forces. 
It started in 1904 \cite{Lamb1904}, and is still studied \cite{Feng2018,Feng2020,Feng2020a} and commonly used as a model problem for developing numerical schemes \cite{Liu2017} today. 

\vspace{5pt}
\noindent a)~\textit{Lamb problem under plain surface}

\begin{figure}[ht!]
\centering
\includegraphics[width=0.9\textwidth]{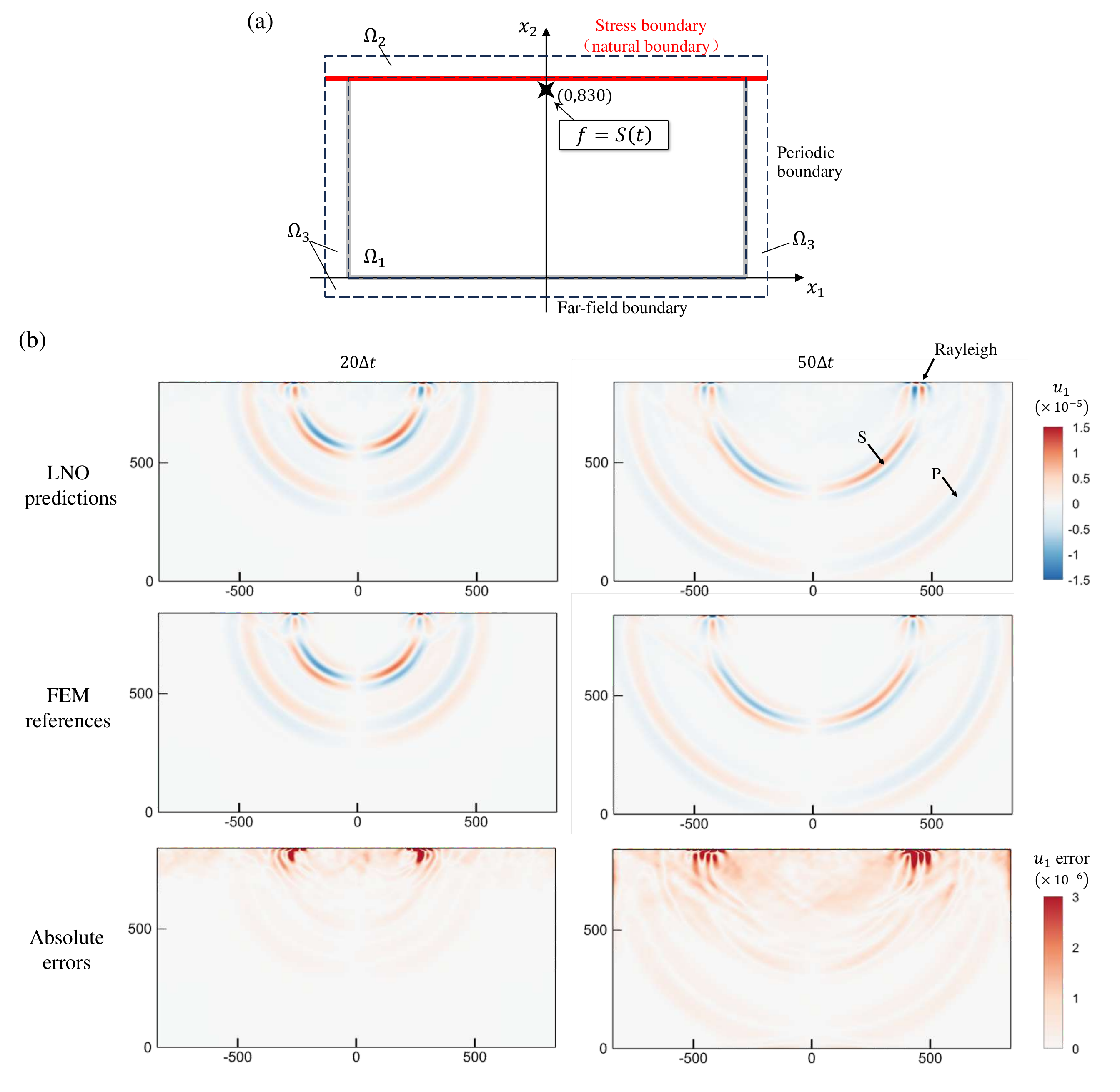}
\caption{The pre-trained LNO solves Lamb problem under a plain surface. 
(a) The model problem and its domain divisions for LNO solving. 
(b) The obtained $u_1$ fields compared to the FEM reference.}
\label{fig:results_lamb_plain}
\end{figure}

We first consider the classic Lamb problem under a plain surface, as in Fig.~\ref{fig:results_lamb_plain}(a). 
The Lamé-Navier equation in Eq.~(\ref{eq:6}) is defined on the domain $\Omega=\left[-840\Delta x,840\Delta x\right]\times\left[0,840\Delta x\right]$ (an approximation of the half-infinite space). 
The upper surface is the natural boundary (stress free with no extra pressure) as
\begin{equation}
\left.\boldsymbol{\sigma}\left(\boldsymbol{x}\right)\cdot\boldsymbol{n}\left(\boldsymbol{x}\right)\right|_{\boldsymbol{x}\in\left[-840,840\right]\times\left\{840\right\}}=\mathbf{0},
\label{eq:18}
\end{equation}
where $\boldsymbol{n}$ is the normal vector of the upper surface, it is constant and equals $\left\{0,1\right\}$ for this case. 
$\boldsymbol{\sigma}$ is the stress tensor with components $\sigma_{ij}$ as
\begin{equation}
\sigma_{ij}=\mu\left(u_{i,j}+u_{j,i}\right)+\lambda u_{k,k}\delta_{ij},\qquad i,j,k=1,2.
\label{eq:19}
\end{equation}
The formula uses the summation convention of tensor. Subscripts after comma denote partial derivatives.  
We use periodic left and right boundaries and far-field bottom boundary to simplify the boundary treatment. 
Following Eq.~(\ref{eq:17}), an external force $S(t), f=8, A=50$ in direction 2 is located at $\boldsymbol{x}=\left\{0,830\Delta x\right\}$ (10 meters under the surface). 
We solve this case by FEM for reference, while the solution is validated by comparing to analytical solution \cite{Feng2018,Feng2020,Feng2020a}. 
See \ref{secA2:2}.

The pre-trained LNO, which is defined in Eq.~(\ref{eq:7}), trained by random samples in section 2.3, and fine-tuned according to section 4.2, starts time marching from $t_0=0.16$ that the source end. Following the workflow in Fig.~\ref{fig:workflow}. 
We first extend and divide the computational domain into $\Omega_1,\ \Omega_2,\ \Omega_3$ as in Fig.~\ref{fig:results_lamb_plain}(a). 
The extending range depends on the LNO model according to the local related range $r_\mathrm{min}$ determined by the local-related condition in Eq.~(\ref{eq:2}) or the LNO architecture as discussed in ~\cite{Ye2023}. 
Then, the pre-trained LNO solves by recurrent time-marching iterations with VDE technique for imposing the natural boundary. 
For each of the time-marching iterations starting from $\boldsymbol{u}_t$ on $\Omega_1$ at the present time level, we first extend $\boldsymbol{u}_t$ to $\Omega_3$ by periodic or far-field extension, which is implemented via periodic or replicate padding. 
Meanwhile, give an initial value to $\boldsymbol{u}_t^\ast$ on $\Omega_2$ to prepare for optimization. 
Then, combine $\boldsymbol{u}_t$ on $\Omega_1\cup\Omega_3$ and $\boldsymbol{u}_t^\ast$ on $\Omega_2$ as the input to the pre-trained LNO to obtain the predicted $\tilde{\boldsymbol{u}}_{t+\Delta t}$ on $\Omega_1$. 
$\tilde{\boldsymbol{u}}_{t+\Delta t}$ satisfies the boundary condition in Eq.~(\ref{eq:18}) by optimizing $\boldsymbol{u}_t^\ast$ on $\Omega_2$. 
Thus, one time-marching step is finished. The complete solution series can be obtained by recurrent time marching following these steps. 
The VDE technique is adopted from \cite{Ye2025}, and its implementation for the present cases is detailed in \ref{secA3}.

The solutions are presented and compared in Fig.~\ref{fig:results_lamb_plain}(b). 
From the displacement fields at two time levels, $10\Delta t$ and $50\Delta t$, it shows that the pre-trained LNO captured the main wave propagating modes correctly, including the compressional wave (P-wave), the shear wave (S-wave), and the Rayleigh wave on the free surface. 
The main characters of these modes, including the speed (reflected in its location at different time levels) and the wavefront shape, are correctly predicted by the pre-trained LNO. 
Correspondingly, errors in most areas of the computational domain are tiny, except near the surface and around the Rayleigh wave. 
This is because of the complex interactions between the elastic wave and the free surface. 
It significantly increases the complexity of the solution, which is much more challenging for LNO to solve. 
Still, the error is under control that, quantitatively, the averaged error at time $50\Delta t$ is $1.06\times{10}^6\, \mathrm{m}$ (0.66\% of the maximal displacement amplitude), which is relatively small. 

\vspace{5pt}
\noindent b)~\textit{Lamb problem under curved surface}

\begin{figure}[ht!]
\centering
\includegraphics[width=0.9\textwidth]{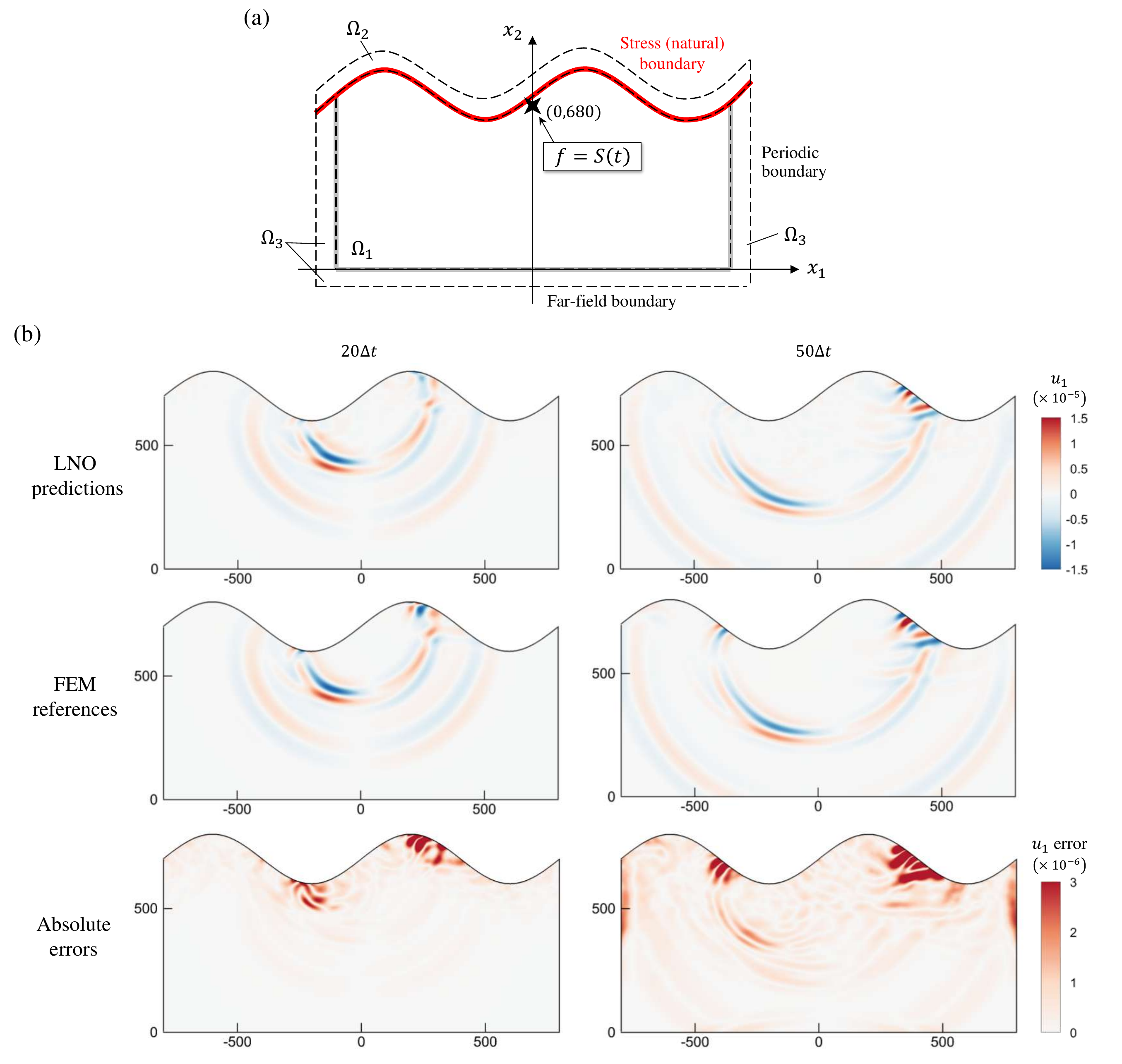}
\caption{The pre-trained LNO solves Lamb problem under a curved surface.
(a) The model problem and its domain divisions for LNO solving. 
(b) The obtained $u_1$ fields compared to the FEM reference.}
\label{fig:results_lamb_curved}
\end{figure}

Next, we consider Lamb problem under a curved surface to further present an application case with curved boundaries and complex Neumann conditions (the natural boundary condition), as in Fig.~\ref{fig:results_lamb_curved}(a). 
The top surface is in a trigonometric shape defined as 
\begin{equation}
x_2=b\left(x_1\right)=100\Delta x\cdot \sin\left({\frac{\pi}{400\Delta x}x_1}\right) +700\Delta x
\label{eq:20}
\end{equation}
The Lamé-Navier equation in Eq.~(\ref{eq:6}) is defined on the domain bounded by $x_1=-800\Delta x$, $x_1=800\Delta x$, $x_2=0$, and $x_2=b\left(x_1\right)$.
Conditions on these boundaries are identical to that for the plain-surface case. 
Note that the normal vector $\boldsymbol{n}$ in Eq.~(\ref{eq:18}) for the natural boundary is no longer constant but becomes $\boldsymbol{n}\left(\boldsymbol{x}\right)=\frac{\left\{-b^\prime\left(x_1\right),1\right\}^\mathrm{T}}{\left\|\left\{-b^\prime\left(x_1\right),1\right\}^\mathrm{T}\right\|}$. 
An external force $S\left(t\right)$, $f=8$, $A=50$ following Eq.~(\ref{eq:17}) in direction 2 is located at $\boldsymbol{x}=\left\{0,680\Delta x\right\}$. 
This case is also solved by FEM for reference.

Identical to that in solving the plain-surface case, we follow the workflow in Fig.~\ref{fig:workflow}, extend and divide the computational domain into $\Omega_1,\ \Omega_2,\ \Omega_3$ as in Fig.~\ref{fig:results_lamb_curved}(a), solve by recurrent time marching using the pre-trained LNO, while the boundaries are imposed by padding (left, right, and bottom) and optimization-based VDE (details can be found in \ref{secA3}).

The solved displacement fields at $10\Delta t$ and $50\Delta t$ are shown and compared in Fig.~\ref{fig:results_lamb_curved}(b). 
The fields are in good agreements. 
The quantitative error at $50\Delta t$ is $1.25\times{10}^{-6}\mathrm{m}$ (0.66\% of the maximal displacement amplitude), which is relatively small. 
In this case, we mainly concern the surface wave propagations (the most destructive part in earthquakes), especially when they pass through the peak (around $x_1=100\Delta x$) and the valley (around $x_1=-100\Delta x$). 
Though the field error in these areas is greater than others in the computational domain, the pre-trained LNO captures qualitative rules correctively that, the wave intensity grows in the peak and weakens in the valley. 
We can also perceive from the fields that there appear dispersed waves in the LNO solutions, which dispersed the energy of the surface wave, thus, it wrongly weakens the wave intensity. 
This issue could be attributed to the boundary treatment. Although it has been discussed and improved \cite{Ye2025}, it is still an interesting open issue.

\vspace{8pt}

Finally, we report the consumed time for the above computation cases. 
In solving the plain-surface Lamb problem, the pre-trained LNO spends 2.455 seconds for one time-marching step, which includes 50 forward inference and backward gradient propagation rounds for the optimization-based boundary imposing. 
In total, it costs 123 seconds to finish the all 50 time marching steps on one NVIDIA Geforce RTX 2080ti GPU. 
For comparison, it costs 5755.5 seconds to obtain the FEM result on one HYGON 7285H CPU, which is about 46 times longer than LNO. 
Similarly, in solving the curved surface Lamb problem, the pre-trained LNO spends 504 seconds for 50 time marching steps, while FEM needs 28981 second to do the same. 
It is about 57 times more than the pre-trained LNO.

\section{Discussions and conclusion}\label{sec6}

This work studied the generalizability of pre-trained LNO and practiced a paradigm for evaluating and improving its accuracy in solving elastodynamic problems. 
Specifically, we modified the classic definition of generalization error for pre-trained LNO (could also be generalized to neural operators) by using scalar features, the amplitude or wavenumber of the input functions, as a representing variable. 
We practiced this definition in investigating the generalization rule of pre-trained LNO in learning viscous Burgers equation and Lamé-Navier equation. 
The observed generalization rule regarding the amplitude is ordinary that, the solving error gradually increases as the amplitude of input functions gets biased from the training ones.
On the contrary, this rule can not be generalized when using the wavenumber to represent the input functions. 
We found that the pre-trained LNO can hardly generalized to input functions with diverse wavenumbers in learning Lamé-Navier equation owing to the diversity shortage of training samples. 
To fix this issue, we supplemented the training data and fine-tuned the pre-trained LNO, thus obtaining a significant improvement in solving elastodynamic problems. 
In solving the benchmark problem of the elastic wave propagation in infinite space, the averaged error on a region of interest is reduced for 90.8\% at most from 32.99\% to 3.04\% of the maximum field value. 

We observed that isolated wavenumber modes of solutions for Lamé-Navier equation can cause the training dataset to lack mode diversity. 
Analyses in this work showed that, this issue appears owing to a special nature of the wave equation. 
The Fourier (trigonometric) modes are isolated during the transient evolution, in other words, the wave equation can not automatically transfer the modes to each other. 
On the one hand, this observation reminded us to pay attention to the mode diversity when generating training data for learning wave-like equations. 
It also suggests that when studying the learning behavior of neural operators, it is necessary to concern about features of the equations to be learned. 
Therefore, more meticulous operator learning studies focusing on representative basic equations (such as the convection equation, diffusion equation, etc.) could be interesting.

Though the generalization of pre-trained LNOs is complex and problem-specific, this work provided a practical route to obtain ‘safe’ solutions. 
The principles concerning the generalizability of LNO are summarized as follows.
\begin{itemize}
\item[$\bullet$] With the assumption of uniqueness of the solution, use scalar features of the input functions for parameterizing the generalization error. 
For feature selection: \textit{amplitude} is certainly needed to be considered; other features should be selected concerning the evolving nature of the equations, for example, \textit{wavenumber} is necessary for wave-like equations such as Lamé-Navier equation.

\item[$\bullet$] Delimit safe ranges regarding each of the selected features. 
Every pre-trained LNO model should be attached with safe ranges regarding the selected features obtained via validating experiments. 

\item[$\bullet$]Check feature values of the solutions obtained from pre-trained LNOs. 
If the feature values are \textit{ever} out of the safe range, the solution becomes unreliable, as the error can accumulate in recurrent time marching in transient computations of dynamic systems. 
\end{itemize}


\appendix

\section{Understanding LNO modeling in learning elastodynamics}\label{secA1}

The analytical solution of linear elastodynamic equation by Green’s function gives a perspective to understand the modeling of LNO. 
Consider an elastic body on domain $V$ with boundary $S$ under distributed time-dependent body force $\boldsymbol{f}\left(\boldsymbol{\xi},\tau\right), \boldsymbol{\xi}\in V,\ \tau>0$, its solution (the time-dependent displacement in direction $n=1,2,3$) is \cite{Zhang2021a}
\begin{equation}
\begin{split}
u_n\left(\boldsymbol{x},t\right)&=\underbrace{\int_{-\infty}^{\infty}{\,\mathrm{d}\tau\iiint_{V}{f_i\left(\boldsymbol{\xi},\tau\right)G_{in}\left(\boldsymbol{\xi},t;\boldsymbol{x},\tau\right)\,\mathrm{d}V\left(\boldsymbol{\xi}\right)}}}_{\left(\mathrm{I}\right)}\\ &+ 
\int_{-\infty}^{\infty}\,\mathrm{d}\tau\oiint_{S}\left\{\underbrace{T_i\left(\boldsymbol{u}\left(\boldsymbol{\xi},\tau\right),\boldsymbol{n}\right)G_{in}\left(\boldsymbol{\xi},t;\boldsymbol{x},\tau\right)}_{\left(\mathrm{II.1}\right)}-\underbrace{n_jC_{jipq}\frac{\partial}{\partial\xi_q}G_{pn}\left(\boldsymbol{\xi},t;\boldsymbol{x},\tau\right)u_i\left(\boldsymbol{\xi},\tau\right)}_{\left(\mathrm{II.2}\right)}\right\}\,\mathrm{d}S\left(\boldsymbol{\xi}\right).
\label{eq:A1}
\end{split}
\end{equation}
Subscripts in this formula represent the tensor components while adopting the summation convention for simplification. $G_{in}$ is the components of $\boldsymbol{G}$, the Green’s function of elastodynamics, representing the displacement induced by a centered unit force pulse. 
Specifically, $G_{in}\left(\boldsymbol{x},t;\boldsymbol{\xi},\tau\right)$ means the displacement in direction $n$ at position $\boldsymbol{x}$ and time $t$ generated by unit point force that acted at position $\boldsymbol{\xi}$ and time $\tau$ in direction $i$. 
$\boldsymbol{T}\left(\boldsymbol{u}\left(\boldsymbol{\xi},\tau\right),\boldsymbol{n}\right)=\boldsymbol{n}\cdot\boldsymbol{\sigma}\left(\boldsymbol{\xi},\tau\right)$ is the inclined plane stress on the boundary. 
$T_i$ are components of $\boldsymbol{T}$. $n_j$ are components of $\boldsymbol{n}$. 
$C_{ijpq}$ are the elastic coefficients. 
Physical meanings of the two integral items in Eq.~(\ref{eq:A1}) are: 
(I) represents the elastic waves generated by the body force; (II) represents the elastic waves generated on the boundaries. 
(II-1) represents the wave generated by ‘release of the stress’ from natural boundaries. 
(II-2) represents the wave ‘reflected’ from rigid boundaries. 
Note that $\boldsymbol{G}$ is specific for cases on different domains with diverse boundary, which brings challenges in practices. 
For infinite space, we have \cite{Zhang2021a}
\begin{equation}
G_{ni}\left(\boldsymbol{x},t;0,0\right)=\frac{3\gamma_n\gamma_i-\delta_{ni}}{4\pi\rho r^3}t\left[H\left(t-t_\mathrm{p}\right)-H\left(t-t_\mathrm{s}\right)\right] + \frac{\gamma_n\gamma_i}{4\pi\rho{v_\mathrm{p}}^2r}\delta\left(t-t_\mathrm{p}\right)-\frac{\gamma_n\gamma_i-\delta_{ni}}{4\pi\rho{v_\mathrm{s}}^2r}\delta\left(t-t_\mathrm{s}\right),
\label{eq:A2}
\end{equation}
where $r=\Vert x\Vert_2=\sqrt{{x_1}^2+{x_2}^2+{x_3}^2}$, $\gamma_n={x_n}/{r}$ is direction coefficient. 
$H\left(t\right)$ is Heaviside function. 
$\delta\left(t\right)$ is delta function while $\delta_{ni}$ is the Kronecker delta.
$t_\mathrm{p}={r}/{v_\mathrm{p}}$ and $t_\mathrm{s}={r}/{v_\mathrm{s}}$ are arrival time of compressional (P) and shear (S) wave, where $v_\mathrm{p}=\sqrt{\left({\lambda+2\mu}\right)/{\rho}}$ and $v_\mathrm{s}=\sqrt{{\mu}/{\rho}}$ are propagating speeds of the two wave modes, respectively. 
$\lambda$ and $\mu$ are the Lamé coefficients.

With the preliminaries, we compare the LNO modeling corresponding to the solution Eq.~(\ref{eq:A1}). 
On the one hand, we consider the away-boundary part solely described by item (I). 
Let $\boldsymbol{f}\left(\boldsymbol{\xi},\tau\right)$ be zero except for $\tau\in\left[t_0-\epsilon_t,t_0\right]$, where $\epsilon_t$ is a small time interval, and with the reciprocity theorem applied on $\boldsymbol{G}$, we have
\begin{equation}
u_n\left(\boldsymbol{x},t\right)=\int_{t_0-\epsilon_t}^{t_0}{\,\mathrm{d}\tau\iiint_{V}{f_i\left(\boldsymbol{\xi},\tau\right)G_{in}\left(\boldsymbol{\xi},t;\boldsymbol{x},\tau\right)\,\mathrm{d}V\left(\boldsymbol{\xi}\right)}}\approx\epsilon_t\iiint_{V}{f_i\left(\boldsymbol{\xi},t_0\right)G_{ni}\left(\boldsymbol{x}-\boldsymbol{\xi},t-t_0;\boldsymbol{0},0\right),\mathrm{d}V\left(\boldsymbol{\xi}\right)}
\label{eq:A3}
\end{equation}
Let $t=t_0$, then $G_{ni}$ is non-zero only when $\boldsymbol{x}\rightarrow\boldsymbol{\xi}$, we have $\boldsymbol{u}\left(\boldsymbol{x},t_0\right)\propto\boldsymbol{f}\left(\boldsymbol{x},t_0\right)$. 
Let $t=t_0+\Delta t,\, \Delta t>0,\, \boldsymbol{u}\left(\boldsymbol{x},t_0+\Delta t\right)$ is unique with respect to $\boldsymbol{f}\left(\boldsymbol{x},t_0\right)$ and can be obtained by substituting Eq.~(\ref{eq:A2}) to Eq.~(\ref{eq:A3}). 
This is the mapping to be learned by LNO from variable $\boldsymbol{u}\left(\boldsymbol{x},t_0\right)\propto\boldsymbol{f}\left(\boldsymbol{x},t_0\right)$ to $\boldsymbol{u}\left(\boldsymbol{x},t_0+\Delta t\right)$.
On the other hand, we consider the near-boundary part described by the item (II.1) and (II.2) in Eq.~(\ref{eq:A1}), which are equations as they are both related to $\boldsymbol{u}\left(\boldsymbol{x},t\right)$. 
The present LNO with VDE technique approximates these two items by constructing virtual displacement fields $\boldsymbol{u}^\ast\left(\boldsymbol{x},t\right)$ on near-boundary domains. 
These virtual fields can be considered induced by a short body force $\boldsymbol{f}^\ast\left(\boldsymbol{x},t\right)$. 
Thus, the mapping in Eq.~(\ref{eq:A3}) approximated by pre-trained LNO is still available within the short time interval $\Delta t$. 
These virtual fields induced by body forces are individual for each time marching step, thus forming a series of virtual displacement fields (short intermittent body forces).

We then discuss features of the mapping from variable $\boldsymbol{u}\left(\boldsymbol{x},t_0\right)\propto\boldsymbol{f}\left(\boldsymbol{x},t_0\right)$ to $\boldsymbol{u}\left(\boldsymbol{x},t_0+\Delta t\right)$ defined in Eq.~(\ref{eq:A3}), specifically about that corresponding to the LNO conception, i.e., the local-related condition and the shifting invariances of time. 
From Eq.~(\ref{eq:A3}) we have
\begin{equation}
\frac{\partial u_n\left(\boldsymbol{x},t\right)}{\partial f_i\left(\boldsymbol{\xi},t_0\right)}=\epsilon_tG_{ni}\left(\boldsymbol{x}-\boldsymbol{\xi},t-t_0;\mathbf{0},0\right).
\label{eq:A4}
\end{equation}
According to Eq.~(\ref{eq:A2}), $G_{ni}\left(\boldsymbol{x},t;\boldsymbol{0},0\right)=0$ when $t<t_\mathrm{p}$. For given $t$, if $\lVert \boldsymbol{x}\rVert_2>v_\mathrm{p}t$, then $t<t_\mathrm{p}$, displacement at $\boldsymbol{x}$ equals zero ($\lVert \boldsymbol{x}\rVert_2=r,t_\mathrm{p}={r}/{v_\mathrm{p}}$). This means that $\boldsymbol{f}\left(\boldsymbol{x},t_0\right)$ only impacts $\boldsymbol{u}\left(\boldsymbol{x},t\right)$ within $\lVert \boldsymbol{x}-\boldsymbol{\xi}\rVert_2>v_\mathrm{p}\left(t-t_0\right)$, i.e., Eq.~(\ref{eq:A2}) equals zero otherwise. 
Assume $\boldsymbol{f}\left(\boldsymbol{x},t_0\right)$ can only impact $\boldsymbol{u}\left(\boldsymbol{x},t_0+\Delta t\right)$ via $\boldsymbol{u}\left(\boldsymbol{x},t_0\right)$, according to the chain rule, we have
\begin{equation}
\frac{\partial u_n\left(\boldsymbol{x},t_0+\Delta t\right)}{\partial f_i\left(\boldsymbol{\xi},t_0\right)}=\sum_m\iiint_V\frac{\partial u_n\left(\boldsymbol{x},t_0+\Delta t\right)}{\partial u_m\left(\boldsymbol{x}_l,t_0\right)}\cdot\frac{\partial u_m\left(\boldsymbol{x}_l,t_0\right)}{\partial f_i\left(\boldsymbol{\xi},t_0\right)}\,\mathrm{d}V\left(\boldsymbol{x}_l\right)
\label{eq:A5}
\end{equation}
The left equals Eq.~(\ref{eq:A4}) for $t=t_0+\Delta t$, it is non-zero within $\lVert \boldsymbol{x}-\boldsymbol{\xi}\rVert_2>v_\mathrm{p}\Delta t$. 
$\frac{\partial u_m\left(\boldsymbol{x}_l,t_0\right)}{\partial f_i\left(\boldsymbol{\xi},t_0\right)}$ on the right equals Eq.~(\ref{eq:A4}) for $t=t_0$, it is non-zero when $\boldsymbol{x}_l\rightarrow\boldsymbol{\xi}$. As $\boldsymbol{u}\left(\boldsymbol{x},t_0\right)\propto\boldsymbol{f}\left(\boldsymbol{x},t_0\right)$, we can infer $\frac{\partial u_n\left(\boldsymbol{x},t_0+\Delta t\right)}{\partial u_m\left(\boldsymbol{x}_l,t_0\right)}$ is non-zero within $\lVert \boldsymbol{x}-\boldsymbol{\xi}\rVert_2>v_\mathrm{p}\Delta t$, i.e., $\boldsymbol{u}\left(\boldsymbol{x},t_0+\Delta t\right)$ only related to $\boldsymbol{u}\left(\boldsymbol{x},t_0\right)$ within a range of $v_\mathrm{p}\Delta t$. 
The related range is finite and small if $\Delta t$ is finite and small, thus confirming the local-related condition in the LNO conception for learning Lamé-Navier equation.
Besides, the shifting invariance of time can be easily explained by Eq.~(\ref{eq:A3}) with the time reciprocity theorem of $\boldsymbol{G}$ that, for variable $t$, when the interval $\Delta t=t-t_0$ is constant, the mapping from variable $\boldsymbol{u}\left(\boldsymbol{x},t_0\right)\propto\boldsymbol{f}\left(\boldsymbol{x},t_0\right)$ to $\boldsymbol{u}\left(\boldsymbol{x},t_0+\Delta t\right)$ will not change.

\section{Reference solutions}\label{secA2}

\subsection{Elastic wave propagation in infinite space}\label{secA2:1}

Analytical solution of this 2D plain-strain elastic wave propagation is given by the displacement integral representation theorem \cite{Zhang2021a} as 
\begin{equation}
u_n^{S(t)}=G_{ni}^{\left(2\right)}\left(\boldsymbol{x},t;\boldsymbol{0},0\right)\ast S(t),\qquad n=1, 2.
\label{eq:B1}
\end{equation}
Here, we have $\boldsymbol{\xi}=\mathbf{0}$ for point source located at the coordinate origin, and $G_{ni}^{\left(2\right)}\left(\boldsymbol{x},t;\boldsymbol{\xi},\tau\right)=G_{ni}^{\left(2\right)}\left(\boldsymbol{x},t-\tau;\boldsymbol{\xi},0\right)$ according to the Betti’s reciprocity theorem. 
The source $S(t)$ is applied in direction $i$. 
The superscript marks Green’s function for 2D plain-strain cases, which can be calculated by integrating the general 3D one in the third dimension as
\begin{equation}
G_{ni}^{\left(2\right)}\left(\boldsymbol{x},t;\boldsymbol{0},0\right)=\int_{-\infty}^{\infty}{G_{ni}\left(\boldsymbol{x}=\left\{x_1,x_2,x_3\right\},t;\boldsymbol{0},0\right)}\,\mathrm{d}x_3,
\label{eq:B2}
\end{equation}
where the 3D Green’s function $G_{ni}\left(\boldsymbol{x},t;\boldsymbol{0},0\right)$ is in Eq.~(\ref{eq:A2}). 
Therefore, the analytical solution is obtained by substituting Eq.~(\ref{eq:A2}) and Eq.~(\ref{eq:B2}) into Eq.~(\ref{eq:B1}). 
The infinite integration is numerically calculated using very fine discretization.

\subsection{Lamb problem}\label{secA2:2}
\begin{figure}[h!]
\centering
\includegraphics[width=0.88\textwidth]{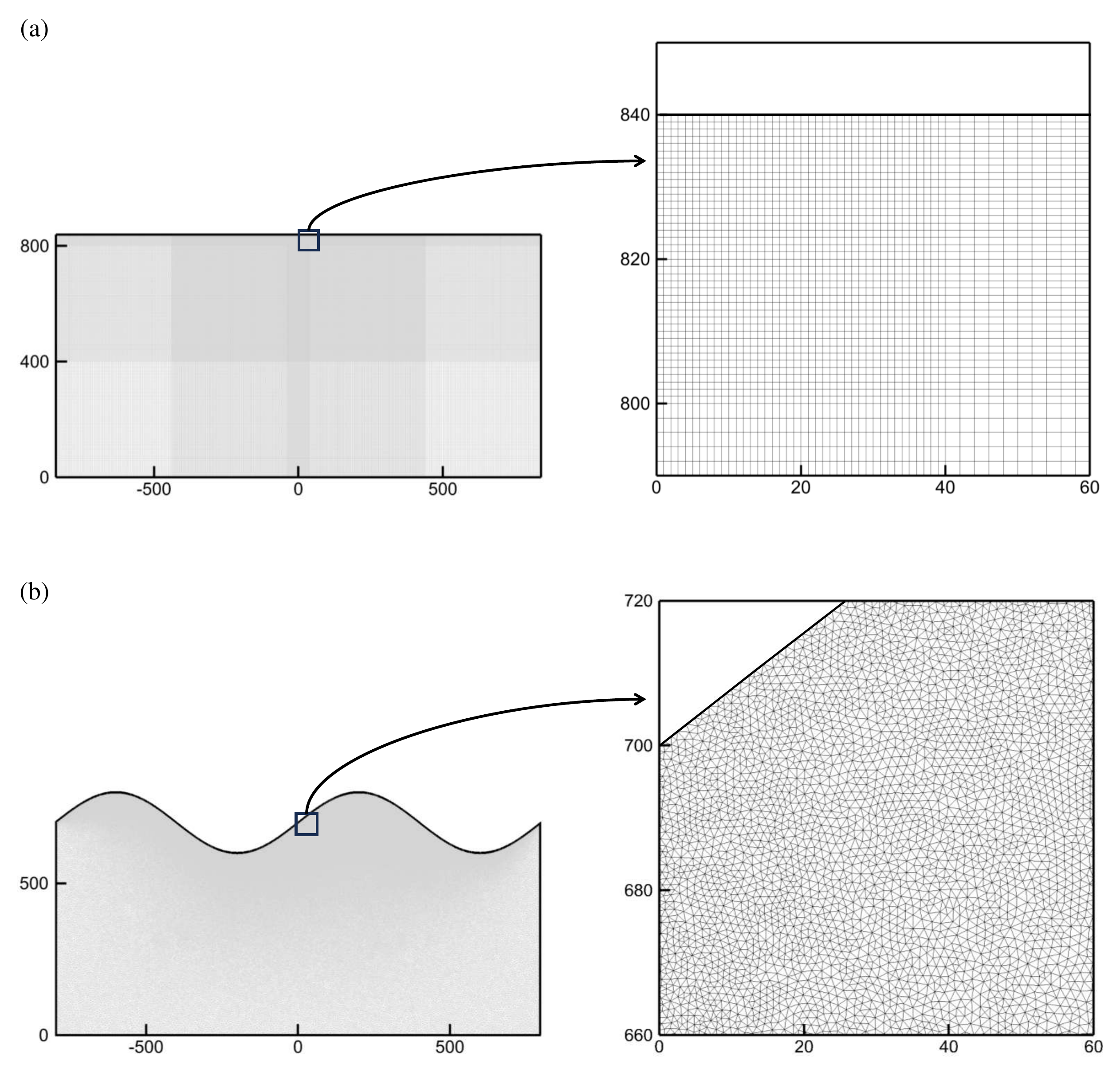}
\caption{Mesh for Lamb problem used in FEM solving. 
(a) Plain surface. 
(b) Curved surface.}
\label{fig:Mesh}
\end{figure}

The reference solution of Lamb problem is calculated numerically by finite element method (FEM). 
The computational domain is discretized as shown in Fig.~\ref{fig:Mesh}. 
For the half-space problem, Cartesian grid is used. 
To reduce the degrees of freedom to be calculated, the grid is arranged to be dense around the source and the free surface with a smallest grid space $\Delta x=1$, and gradually transit to coarse ones in other regions with a largest grid space $\Delta x=4$. 
The total number of the nodes is 232221. 
Unstructured triangular elements are used to solve the curved surface problem. 
The minimal grid space is set around $\Delta x=1$. 
The total number of nodes is 281968. 
The implicit Newmark $\beta$ method is chosen for temporal discretization, which is unconditionally stable when the artificial parameters are properly chosen \cite{Belytschko2014}. 
The discretized algebraic system is solved iteratively by Newton-Raphson method.
The time interval $\Delta t=2\times10^{-4}$.

\begin{figure}[ht!]
\centering
\vspace{10pt}
\includegraphics[width=0.78\textwidth]{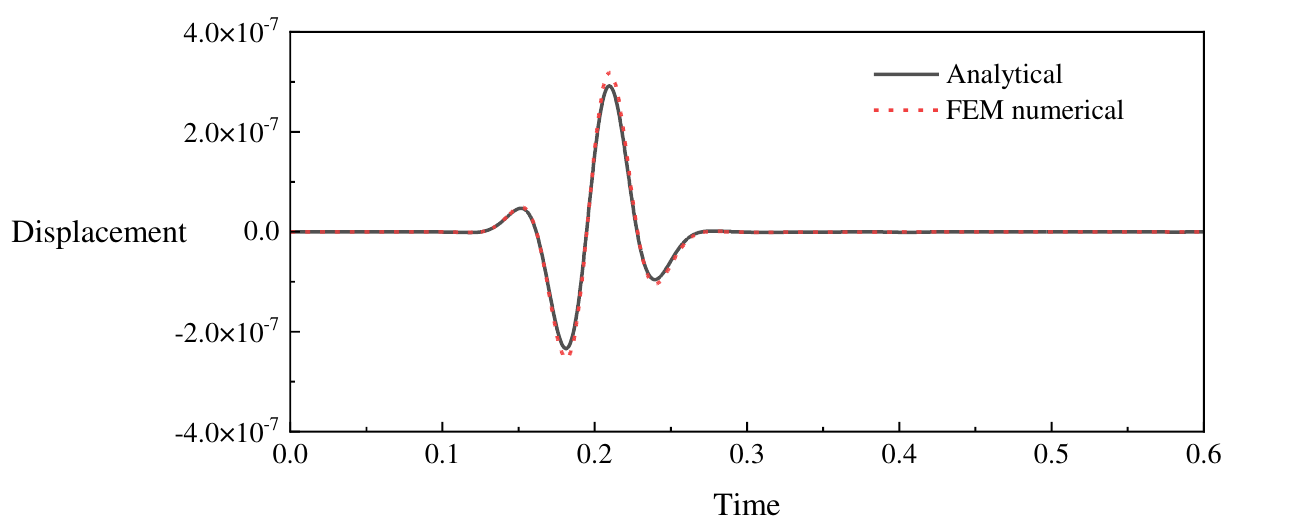}
\caption{Validating the numerical reference solution referring to analytical result for Lamb porblem.}
\label{fig:Reference_validation}
\end{figure}

To validate the FEM solution, the result is compared with the analytical solution in the half-space problem~\cite{Feng2018}. 
The analytical solution is computed by numerically integrating the 3D analytical solution along the third axis. 
Fig.~\ref{fig:Reference_validation} shows the time history of the vertical displacement $u_2$ at a monitor point on the free surface $\boldsymbol{x}=\left\{50,840\right\}$.
A slight discrepancy is observed because in the practice of FEM, the pulse source term is approximated by a continuous Gaussian distribution to fit the spatial discretization. 
Overall, the FEM result mostly matches well with the analytical solution. 

\section{Virtual boundary extension for elastodynamic problems}\label{secA3}

This appendix introduces the technique, virtual domain extension (VDE) by gradient descent-based optimization \cite{Ye2025}, adopted here for elastodynamic problems. 

\subsection{Conception and method}\label{secA3:1}

\begin{figure}[htbp]
\centering
\vspace{10pt}
\includegraphics[width=\textwidth]{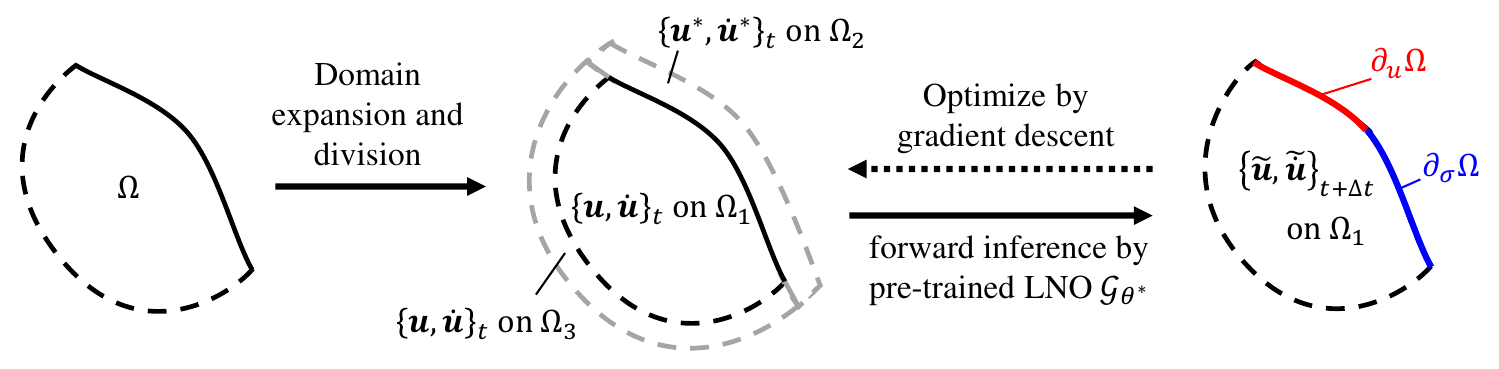}
\caption{Optimize-based virtual domain extension for elastodynamic problems.
Dotted and solid lines are respectively artificial and physical boundaries on the computational domain $\Omega$ of a general demo problem. $\partial_u\Omega$ is the displacement boundary (Dirichlet). $\partial_\sigma\Omega$ is the stress/pressure boundary (Neumann).
These two kinds of boundaries compose the complete physical boundary.}
\label{fig:VDE_conception}
\end{figure}

The primary idea of VDE technique is to construct virtual fields $\boldsymbol{u}_t^\ast$ and ${\dot{\boldsymbol{u}}}_t^\ast$ on an extended domain outward of the physical boundaries, which makes the output fields satisfy the corresponding boundary value conditions. 
The overall conception is demonstrated in Fig.~\ref{fig:VDE_conception}. 
For one time-marching step, the LNO model receives $\boldsymbol{u}_t$ and ${\dot{\boldsymbol{u}}}_t$ on $\Omega_1\cup\Omega_3$ combined with $\boldsymbol{u}_t^\ast$ and ${\dot{\boldsymbol{u}}}_t^\ast$ on the extended $\Omega_2$ as input, then, it outputs $\tilde{\boldsymbol{u}}_{t+\Delta t}$ and $\tilde{\dot{\boldsymbol{u}}}_{t+\Delta t}$ on $\Omega_1$. 
VDE makes the outputs satisfy the physical boundaries by gradient descent-based optimization, which is detailed as follows.

Consider an elastodynamic problem defined on a general domain $\Omega$ with both artificial and physical boundaries. 
The physical boundaries are comprised of a displacement part $\partial_u\Omega$ and a stress part $\partial_\sigma\Omega$. 
For displacement boundary, it is constrained by
\begin{equation}
\boldsymbol{u}\left(\boldsymbol{x}_u,t\right)=\bar{\boldsymbol{u}}\left(\boldsymbol{x}_u,t\right),\qquad\dot{\boldsymbol{u}}\left(\boldsymbol{x}_u,t\right)=\dot{\bar{\boldsymbol{u}}}\left(\boldsymbol{x}_u,t\right),\qquad \boldsymbol{x}_u\in\partial_u\Omega,\,t>0,
\label{eq:C1}
\end{equation}
where $\bar{\boldsymbol{u}}$ is the given displacement constrain and $\dot{\bar{\boldsymbol{u}}}$ is the given velocity constrain on the boundary points. 
It is required that $\dot{\bar{\boldsymbol{u}}}$ is the time derivative of $\bar{\boldsymbol{u}}$, for static boundary displacement, we have $\dot{\bar{\boldsymbol{u}}}=\boldsymbol{0}$. 
The stress boundary is constrained by
\begin{equation}
\boldsymbol{\sigma}\left(\boldsymbol{x}_\sigma,t\right)\cdot\boldsymbol{n}\left(\boldsymbol{x}_\sigma\right)=\bar{\boldsymbol{\sigma}}\left(\boldsymbol{x}_\sigma,t\right),\qquad\boldsymbol{x}_\sigma\in\partial_\sigma\Omega,\,t>0.
\label{eq:C2}
\end{equation}
The stress components in $\boldsymbol{\sigma}$ are calculated from the displacement $\boldsymbol{u}$ by Eq.~(\ref{eq:19}). 
$\boldsymbol{n}$ is the normal vector outward the boundary.

These boundary conditions are satisfied, i.e., the output $\tilde{\boldsymbol{u}}_{t+\Delta t}$ satisfies Eq.~(\ref{eq:C1}) and Eq.~(\ref{eq:C2}), by solving the following optimization problem
\begin{equation}
\min_{\left\{\boldsymbol{u}^\ast,{\dot{\boldsymbol{u}}}^\ast\right\}_t\left(\boldsymbol{x}_2\right),\,\boldsymbol{x}_2\in\Omega_2}{\mathcal{L}_u+\mathcal{L}_\sigma},
\label{eq:C3}
\end{equation}
where
\begin{equation}
\begin{split}
\mathcal{L}_u&=\int_{\partial_u\Omega}{\lVert\tilde{\boldsymbol{u}}_{t+\Delta t}\left(\boldsymbol{x}_u\right)-\bar{\boldsymbol{u}}\left(\boldsymbol{x}_u,t+\Delta t\right)\rVert} +{\lVert\tilde{\dot{\boldsymbol{u}}}_{t+\Delta t}\left(\boldsymbol{x}_u\right)-\dot{\bar{\boldsymbol{u}}}\left(\boldsymbol{x}_u,t+\Delta t\right)\rVert}\,\mathrm{d}\boldsymbol{x}_u, \\
\mathcal{L}_\sigma&=\int_{\partial_\sigma\Omega}{\lVert\tilde{\boldsymbol{\sigma}}_{t+\Delta t}\left(\boldsymbol{x}_\sigma\right)\cdot\boldsymbol{n}\left(\boldsymbol{x}_\sigma\right)-\bar{\boldsymbol{\sigma}}\left(\boldsymbol{x}_\sigma,t+\Delta t\right)\rVert}\,\mathrm{d}\boldsymbol{x}_\sigma. 
\label{eq:C4}
\end{split}
\end{equation}
The $\tilde{\boldsymbol{u}}_{t+\Delta t}$, $\tilde{\dot{\boldsymbol{u}}}_{t+\Delta t}$ are given by the pre-trained LNO $\mathcal{G}_{\theta^\ast}$ as
\begin{equation}
\left\{\tilde{\boldsymbol{u}},\tilde{\dot{\boldsymbol{u}}}\right\}=\mathcal{G}_{\theta^\ast}\left(\left\{\left\{\boldsymbol{u},\dot{\boldsymbol{u}}\right\}_t\left(\boldsymbol{x}_1\right),\left\{\boldsymbol{u}^*,\dot{\boldsymbol{u}}^*\right\}_t\left(\boldsymbol{x}_2\right)|\boldsymbol{x}_1\in\Omega_1\cup\Omega_3,\boldsymbol{x}_2\in\Omega_2\right\}\right),\qquad\boldsymbol{x}\in\Omega_1.
\label{eq:C5}
\end{equation}
The $\tilde{\boldsymbol{\sigma}}_{t+\Delta t}$ can be calculated from $\tilde{\boldsymbol{u}}_{t+\Delta t}$ by Eq.~(\ref{eq:19}). 
Noticing that these computations are all continuous and differentiable, including that in the pre-trained LNO $\mathcal{G}_{\theta^\ast}$ according to literature \cite{Ye2025}. Writing $\mathcal{L}_\mathcal{B}=\mathcal{L}_u+\mathcal{L}_\sigma$, the derivative $\frac{\partial\mathcal{L}_\mathcal{B}}{\partial\boldsymbol{u}_t^\ast\left(\boldsymbol{x}_2\right)}$ with respect to the field values to be constructed in VDE exists. 
Thus, it can be calculated by back-propagation \cite{Lecun1998}, and the optimization problem can be approximately solved by optimizers such as the Adam \cite{Kingma2015}. 

\subsection{Implementation on Lamb problems}\label{secA3:2}

The optimization problem defined in Eq.~(\ref{eq:C3}) has a big variable space of $\left\{\boldsymbol{u}^\ast,{\dot{\boldsymbol{u}}}^\ast\right\}_t\left(\boldsymbol{x}_2\right),\boldsymbol{x}_2\in\Omega_2$, in discretized case, it is in $\mathbb{R}^{4\times N_{\Omega_2}}$ as there are four fields $\left\{u_1^\ast,u_2^\ast,{\dot{u}}_1^\ast,{\dot{u}}_2^\ast\right\}$ and $N_{\Omega_2}$ is the number of space points in $\Omega_2$. 
Such big variable space is usually over-flexible and requires regularization to avoid oscillations. 
For Lamb problems in this work, we use two measures for regularization. One is, we did not use the displacements $\left\{u_1^\ast,u_2^\ast\right\}$ as optimizing variables and set them by reflection along the stress-free boundary. 
This measure is taken because reflection makes zero spatial derivatives of the displacement in the normal direction that ${\partial\boldsymbol{u}}/{\partial\boldsymbol{n}}=\mathbf{0}$, resulting in zero stress on the boundary which satisfies the stress-free condition. 
The other measure is, we applied mean filtering using square window of $5\times5$ on the velocity fields $\left\{{\dot{u}}_1^\ast,{\dot{u}}_2^\ast\right\}$ before sending into the LNO model. 

We obtain for field values on the curved boundary by interpolation in the case of Lamb problem under curved surface. 
The field values $U$ on curves at point $\boldsymbol{X}_j$ are
\begin{equation}
U\left(\boldsymbol{X}_j\right)=\sum_{i=1}^{N_x}u\left(\boldsymbol{x}_i\right)\delta_h\left(\boldsymbol{x}_i-\boldsymbol{X}_j\right){\Delta x}^2,\qquad j=1,2,\ldots,N_X,
\label{eq:C6}
\end{equation}
where $N_x$ is the number of related points on Cartesian grad of size $\Delta x$. 
$N_X$ is the number of points on the curved boundary. 
$\delta_h$ is the approximated four-point delta function \cite{Peskin2002} that
\begin{equation}
\delta_h\left(\boldsymbol{x}-\boldsymbol{X}\right)=\frac{1}{{\Delta x}^2}d\left(\frac{\boldsymbol{x}-\boldsymbol{X}}{\Delta x}\right).
\label{eq:C7}
\end{equation}
For 2D cases
\begin{equation}
d\left(\boldsymbol{r}\right)=w\left(r_1\right)w\left(r_2\right),\qquad \boldsymbol{r}=\left\{r_1,r_2\right\},
\label{eq:C8}
\end{equation}
where
\begin{equation}
w\left(r_i\right)=\left\{
\begin{aligned}
\frac{1}{8}\left(3-2\lvert r_i\rvert+\sqrt{1+4\lvert r_i\rvert-4\lvert r_i\rvert^2}\right),\qquad\qquad\,&\lvert r_i\rvert<1\\
\frac{1}{8}\left(5-2\lvert r_i\rvert-\sqrt{-7+12\lvert r_i\rvert-4\lvert r_i\rvert^2}\right),\qquad1\leq&\lvert r_i\rvert<2\\
~0,\qquad\qquad\qquad\quad\qquad\qquad&\lvert r_i\rvert>2
\end{aligned}
\right.\quad.
\label{eq:C9}
\end{equation}
Thus, we obtained the field values (displacement or stress) on curved boundary and the optimization objective defined in Eq.~(\ref{eq:C3}) can be implemented for imposing the boundary condition.

\begin{figure}[htb]
\centering
\includegraphics[width=0.7\textwidth]{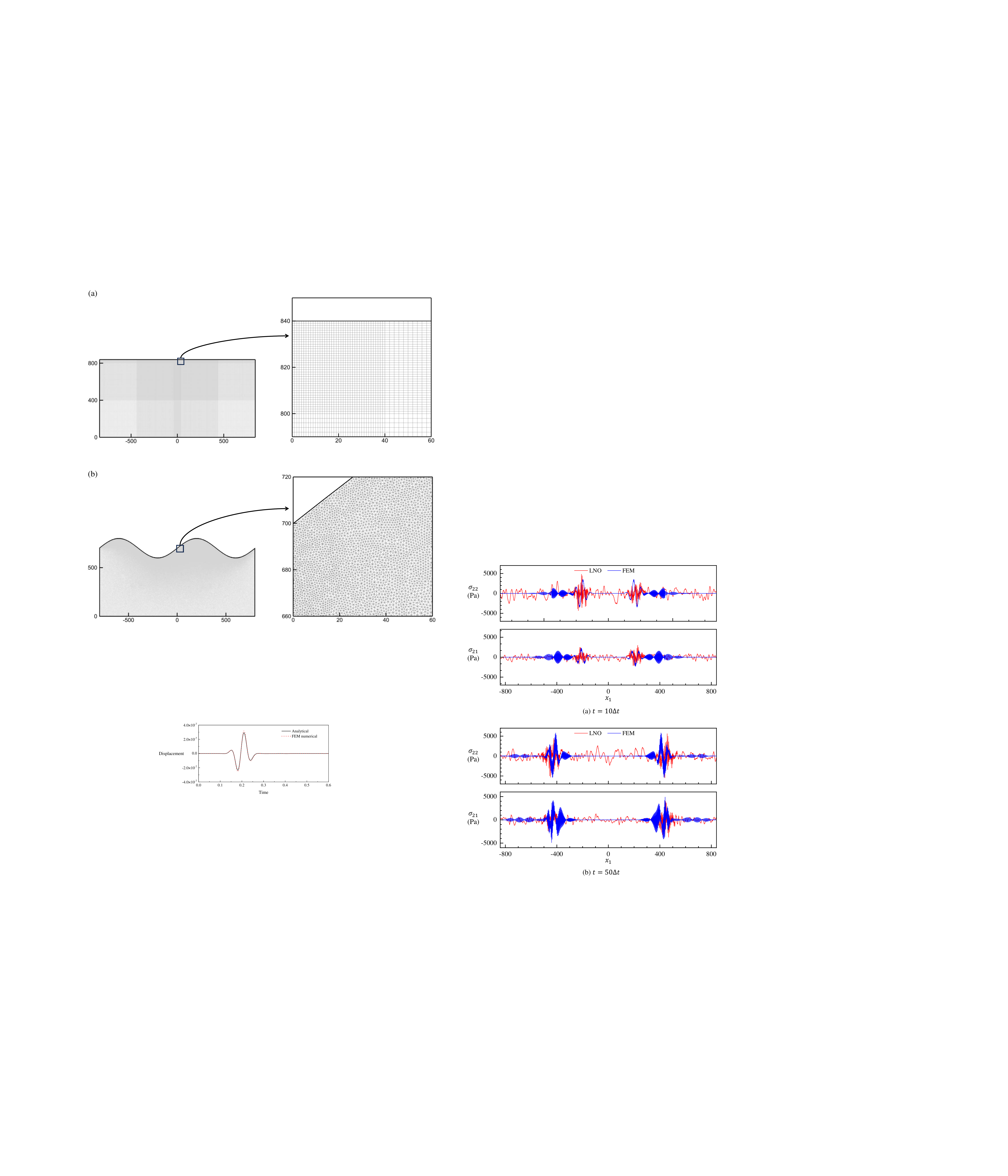}
\caption{Boundary value residuals (estimated by finite difference method) at two time levels when solving Lamb problem under a plain surface. 
(a) $t=10\Delta t$. 
(b) $t=50\Delta t$.}
\label{fig:boundary_res}
\end{figure}

We present the optimized boundary values in solving Lamb problem under a plain surface in Fig.~\ref{fig:boundary_res}. 
The boundary value residuals are estimated by the finite difference method and compared to that in the FEM reference. 
It is seen that the present optimization-based VDE finely controls the residuals during LNO solving, where the residual values are relatively small and comparable with that of the FEM solver. 
Note that the natural boundaries are satisfied via Galerkin variation in FEM in a weak form, thus it also lays relatively small residuals when estimating by the finite difference method.

\bibliographystyle{elsarticle-num} 
\bibliography{library.bib}









\end{document}